\documentclass[aps,showpacs,twocolumn,nofootinbib,prd,superscriptaddress]{revtex4}

\usepackage{amsmath}
\usepackage{latexsym}
\usepackage{graphicx}
\usepackage[usenames]{color}
\usepackage{ulem}

\newcommand{\Rext}{r_\mathrm{ext}}

\newcommand{\Mf}{M_{\rm f}}
\newcommand{\MAH}{M_{\rm AH}}
\newcommand{\MADM}{M_{\rm ADM}}
\newcommand{\Mirr}{M_{\rm irr}}

\newcommand{\MSun}{M_{\odot}}
\newcommand{\afunc}{P}

\newcommand{\Of}{\Omega_{\rm f}}
\newcommand{\omQ}{\omega_{\rm QNM}}

\newcommand{\Odotmax}{\dot{\Omega}_0}

\newcommand{\etal}{\emph{et al.} \,}
\newcommand{\Yslm}{\,_{-2}Y_{\ell}^{m}}
\newcommand{\atil}{\tilde{\alpha}}

\newcommand{\beq}{\begin{equation}}
\newcommand{\eeq}{\end{equation}}
\newcommand{\bea}{\begin{eqnarray}}
\newcommand{\eea}{\end{eqnarray}}
\newcommand{\ba}{\begin{array}}
\newcommand{\ea}{\end{array}}

\begin{document}
\normalem

\title{Mergers of black-hole binaries with aligned spins: Waveform characteristics}

\author{Bernard J. Kelly}
\affiliation{CRESST \& Gravitational Astrophysics Laboratory, NASA/GSFC, 8800 Greenbelt Rd., Greenbelt, MD 20771, USA}
\affiliation{Department of Physics, University of Maryland, Baltimore County, 1000 Hilltop Circle, Baltimore, MD 21250, USA}
\author{John G. Baker}
\affiliation{Gravitational Astrophysics Laboratory, NASA Goddard Space Flight Center, 8800 Greenbelt Rd., Greenbelt, MD 20771, USA}
\author{William D. Boggs}
\affiliation{Department of Physics, University of Maryland, College Park, MD 20742, USA}
\author{Sean T. McWilliams}
\affiliation{Institute for Strings, Cosmology and Astroparticle Physics (ISCAP), Columbia University, New York, NY 10027, USA}
\affiliation{Department of Physics, Princeton University, Princeton, NJ 08544, USA}
\author{Joan Centrella}
\affiliation{Gravitational Astrophysics Laboratory, NASA Goddard Space Flight Center, 8800 Greenbelt Rd., Greenbelt, MD 20771, USA}

\date{\today}

\begin{abstract}
We conduct a descriptive analysis of the multipolar structure of gravitational-radiation waveforms from 
equal-mass aligned-spin mergers, following an approach first presented in the complementary context of nonspinning
black holes of varying mass ratio [Baker \etal Phys. Rev. D \textbf{78}, 044046 (2008)].
We find that, as with the nonspinning mergers, the dominant waveform mode 
phases evolve together in lock-step through inspiral and merger, supporting the previous waveform description 
in terms of an adiabatically rigid rotator driving gravitational-wave emission -- an \emph{implicit
rotating source} (IRS). We further apply the late-time merger-ringdown model for the \emph{rotational frequency}
introduced in Baker \etal (2008), along with an improved amplitude model appropriate for the dominant
$(2,\pm 2)$ modes. This provides a quantitative description of the merger-ringdown waveforms, and suggests that the
major features of these waveforms can be described with reference only to the intrinsic parameters associated with
the state of the final black hole formed in the merger. 
We provide an explicit model for the merger-ringdown radiation, and demonstrate that this model
agrees to fitting factors better than 95\% with the original numerical waveforms for system masses 
above $\sim 150 \MSun$. This model may be directly applicable to gravitational-wave detection of intermediate-mass
black-hole mergers.
\end{abstract}


\maketitle

\section{Introduction}
\label{sec:intro}

Black-hole-binary mergers are a key target of ground-based and space-based 
gravitational-wave observations.  The strongest radiation is produced just as
the two black holes join to become one, and can only be fully understood
through explicit numerical simulations.
Since the first stable evolutions of black-hole-binary mergers
\cite{Pretorius:2005gq,Pretorius:2006tp,Baker:2005vv,Campanelli:2005dd}, and after it was
established that the gravitational waveforms from these evolutions were universal, and
consistent across codes and methodologies \cite{Baker:2006yw,Baker:2007fb,Hannam:2009hh},
researchers have turned their attention to how the results of numerical relativity can most
usefully be supplied to the gravitational-wave data-analysis community.

After studying the equal-mass nonspinning case, researchers have had to address the complexity
problem of more generic systems. Even allowing for simple scaling by total mass $M = M_1 + M_2$,
and assuming zero eccentricity, such systems span a seven-dimensional parameter space:
$\{\eta, \vec{S_1}, \vec{S_2}\}$, where $\eta = M_1 M_2/M^2$ is the \emph{symmetric mass ratio} of the
binary, and $\vec{S_i}$ is the spin angular momentum vector of hole $i$. 

Early surveys of the waveform parameter space have restricted themselves to the $\eta$-dependence of
nonspinning systems. In \cite{Baker:2008mj}, the authors investigated the multipole
structure of merger waveforms from such systems, noting that the strongest subdominant modes shared many
characteristics with the dominant quadrupole, and that they could be collectively described by an
\emph{implicit rotating source} model of the binary. The authors used this observation to construct a multi-mode
gravitational-wave template family for such binary systems, as an alternative to more usual effective-one-body (EOB)
templates \cite{Buonanno:2007pf,Buonanno:2009qa}.

While we may assume
that $\eta$ and $|\vec{S_i}|$ remain essentially constant throughout inspiral and merger, the spin
directions generally evolve, so a useful parametrization of the system should take care to distinguish
components of the spin-direction space with physically distinct effects on the waveforms \cite{Galley:2010rc}.

An obvious cut in parameter space to consider is that of spins aligned (or anti-aligned) with the
orbital angular momentum. These systems will not precess, but exhibit observationally significant 
spin-orbit effects, distinguishing them from nonspinning binaries in their dynamics and resulting waveforms \cite{Campanelli:2006uy}.
High-accuracy waveforms from such evolutions have been produced and studied by several groups
\cite{Shoemaker:2008pe,Reisswig:2009vc,Chu:2009md,Mosta:2009rr}.
Such systems have been partially characterized by \cite{Hannam:2010ec}, using a variant of the frequency
model from \cite{Baker:2008mj}. The frequency-domain phenomenological templates of Ajith \etal have been extended
to cover both mass ratio and total aligned spin \cite{Ajith:2009bn,Santamaria:2010yb}, at least for the dominant
modes, and attempts have been made to extend these to more generic systems \cite{Sturani:2010ju,Sturani:2010yv}.

A key result from our investigation of the dominant modes of nonspinning unequal-mass binary waveforms
\cite{Baker:2008mj} was that these modes had phases that evolved together in lock-step through inspiral
and merger. This agreement was especially impressive for the $\ell=m$ modes, leading to the development of a
heuristic picture of the binary system as a  rigid rotator (at least in the adiabatic limit) driving
gravitational-wave emission. We dubbed this the \emph{implicit rotating source} (IRS) picture.

A secondary result of this picture was the possibility of developing a simple model for the
time-development of the dominant and leading subdominant modal frequencies $\omega_{\ell m}$
in terms of a single \emph{rotational frequency} $\Omega(t)$:
\[
\omega_{\ell m} = m \Omega(t).
\]
We also presented a simple model for the corresponding mode amplitudes, leading to the possibility of
a new approach to time-based gravitational-waveform templates. In fact, we developed such a template
proposal, the IRS-EOB templates, as an alternative to the effective-one-body templates of
\cite{Buonanno:2007pf,Buonanno:2009qa}, which terminate the signals by matching to a superposition of
quasinormal-mode (QNM) frequencies.

In this paper, we look at the dominant waveform modes from some aligned-spin systems, and ask the
following general questions: Does the general IRS picture still hold? How do the features of
aligned-spin mergers compare with those of nonspinning mergers? Can we quantify the main features of the
merger ringdown with a simple analytic model?

The rest of this paper is laid out as follows. In Sec.~\ref{sec:sims}, we introduce the binary
systems studied and the numerical methods used to simulate them. In Sec.~\ref{sec:radiation}, we present
results for the final black-hole states and an IRS descriptive characterization of the waveforms
from our numerical evolutions. In Sec.~\ref{sec:model}, we analyze the late portions of these
waveforms in more detail, and apply the analytic modeling approach of \cite{Baker:2008mj} to the
dominant-mode frequencies and (with improvements) to the amplitude model, concluding with an explicit
parametrization approximating the $(2,2)$ results of all our simulations. 
In Sec.~\ref{sec:match}, we investigate the
quality of the new models compared to the numerical waveforms in the context of the Advanced LIGO
detector. We conclude with some discussion in Sec.~\ref{sec:discuss}. Some extra detail on the
convergence of the numerical simulations is given in the Appendix.

\section{Simulations}
\label{sec:sims}

To investigate the nature of aligned-spin binary waveforms, we carried out a series of numerical
evolutions for equal-mass systems with zero spin (\texttt{X1\_00}), spins aligned with the initial
orbital angular momentum (\texttt{X1\_UU}), anti-aligned (\texttt{X1\_DD}), or mixed (\texttt{X1\_UD}).
We also re-ran, for purposes of comparison, the case of a 4:1 nonspinning binary (\texttt{X4\_00}).

\begin{table*}
  \caption{Physical and numerical parameters of the initial data for all the runs
  presented. $m_{1,p}$ and $m_{2,p}$ are the bare puncture masses of the two
  pre-merger holes. $r_0$ and $P_0$ are the initial coordinate separation and
  (transverse) linear momentum, respectively, giving rise to a total initial orbital
  angular momentum $L_0$. $\MADM$ is the total energy of the initial data. The total
  infinite-separation total mass $M$ of the system is estimated by $\MAH$, the
  sum of the initial (apparent) horizon masses of the two holes, calculated at $t = 100$.
  finally, we quote the approximate observed eccentricity \eqref{eq:ecc_def}.}
\begin{tabular}{c rrrrrrrrrrr}
\hline \hline
run name        & $m_{1,p} = m_{2,p}$ & $S_{1z}$  & $S_{2z}$  & $r_0$   & $P_{0t} (\times 10^2)$ & $P_{0r} (\times 10^4)$ & $L_0$    & $\MADM$  & $\sum_i M_{{\rm ADM}, i}$ & $\MAH$   & $e_{\Omega, \rm max}$\\
\hline
\texttt{X1\_00} & 0.4872312           &  0.0      &  0.0      & 11.0000 & 9.00993                & 7.09412                & 0.991092 & 0.990514 & 1.000050                  & 1.000050 & 0.002 \\
\texttt{X1\_UU} & 0.301805            &  0.2      &  0.2      &  8.2013 & 10.3248                & 0.0                    & 0.846768 & 0.988459 & 1.000908                  & 1.000550 & 0.01 \\
\texttt{X1\_DD} & 0.390411            & -0.159125 & -0.159125 & 11.9837 & 8.83600                & 1.20000                & 1.058879 & 0.990453 & 0.998794                  & 0.998686 & 0.01 \\
\texttt{X1\_UD} & 0.301805            &  0.2      & -0.2      & 11.0000 & 9.00993                & 7.09412                & 0.991092 & 0.990024 & 0.999222                  & 0.998834 & 0.002 \\
\texttt{X4\_00} & 0.7900, 0.1890      &  0.0      &  0.0      &  8.4702 & 6.95662                & 0.0                    & 0.589240 & 0.992912 & 1.000310                  & 1.000315 & 0.02 \\
\hline \hline
\end{tabular}
\label{table:RunData}
\end{table*}
%
The physical parameters of these evolutions are presented in Table~\ref{table:RunData}. The initial
momenta of the equal-mass binaries, with the exception of \texttt{X1\_UU}, were chosen by integrating
the post-Newtonian equations of motion, as outlined in \cite{Husa:2007rh,Campanelli:2008nk}, with spin
contributions to the Hamiltonian adapted from \cite{Buonanno:2005xu,Damour:2007nc,Porto:2006bt,Steinhoff:2007mb,Steinhoff:2008ji}
(although we work in the Arnowitt-Deser-Misner (ADM) gauge, the results from harmonic gauge using effective-field theory
\cite{Porto:2006bt} have been shown to be equivalent \cite{Porto:2007tt,Hergt:2010pa}),
and the flux from \cite{Blanchet:2006gy}. For the \texttt{X1\_UU} configuration, we used simpler quasicircular
initial parameters with no initial ingoing radial momentum. For the \texttt{X4\_00} data, we retained the quasicircular
initial parameters used in \cite{Baker:2008mj}.

The equal-mass runs were carried out with our \textsc{Hahndol} evolution code \cite{Imbiriba:2004tp} using the
{\sc Paramesh} mesh-refinement infrastructure \cite{MacNeice00}. The new \texttt{X4\_00} data, however,
use {\sc Hahndol} paired with the Einstein Toolkit \cite{etk_web} release of the Cactus Computational
Toolkit \cite{cactus_web} and the {\sc Carpet} mesh-refinement driver \cite{carpet_web}.

In all cases, the initial data are of the standard Brandt-Br\"{u}gmann type \cite{Brandt:1997tf}, using
the Bowen-York \cite{Bowen:1980yu} prescription for extrinsic curvature that exactly satisfies the momentum constraint.
We solve the remaining Hamiltonian constraint using the {\sc TwoPunctures} spectral code \cite{Ansorg:2004ds}.
This code also supplies the total ADM energy $\MADM$ of the system, as well as the individual
``puncture ADM masses'' $M_{{\rm ADM}, i}$, to very high precision. We note, however, that for highly spinning
or boosted Bowen-York-type data, a measurable amount of radiation energy may be included in these puncture ADM
masses, but then escape to infinity \cite{Gleiser:1997ng,Campanelli:2005ia,Choi:2007eu,Hannam:2009hh}; thus
the initial puncture ADM mass may not be the optimal measure
of pre-merger black-hole mass. These quantities are also listed in Table~\ref{table:RunData}.

To evolve these initial data, we employ the BSSNOK 3+1 decomposition of Einstein's vacuum equations
\cite{Nakamura:1987zz,Shibata:1995we,Baumgarte:1998te}, with the alternative conformal variable suggested in
\cite{vanMeter:2006g2n,Tichy:2007hk,Marronetti:2007wz}, constraint-damping terms suggested in \cite{Duez:2004uh},
and the dissipation terms suggested in \cite{Kreiss73,Hubner:1999th}.
Our gauge conditions are the specific 1+log lapse and Gamma-driver shift described in \cite{vanMeter:2006vi},
which constitute a variant of the now-standard ``moving punctures'' approach \cite{Campanelli:2005dd,Baker:2005vv}.

\begin{table*}
  \caption{Initial grid structure of the different simulations. The leftmost number is the outer extent
          of the Cartesian grid, with resolution doubled (grid spacing halved) within each new refinement
          level.}
\begin{tabular}{c rr}
\hline \hline
run name        & outer (fixed) grid structure & inner (moving) grid structure \\
\hline
\texttt{X1\_00}, \texttt{X1\_DD}, \texttt{X1\_UD} & [1536,768,384,192,144,72,24,12,8] & [3.0,1.5,0.75] \\
\texttt{X1\_UU}                    & [1536,768,384,192,96,72,24,12,8] & [3.0,1.5,0.75] \\
\texttt{X4\_00} (larger puncture)  & [2048,1024,512,256,160,96] & [20, 10, 5, 2.75, 1.5] \\
\texttt{X4\_00} (smaller puncture) & [2048,1024,512,256,160,96] & [20, 10, 5, 2.5, 1.25, 0.6875, 0.375] \\
\hline \hline
\end{tabular}
\label{table:GridData}
\end{table*}

The four equal-mass simulations -- \texttt{X1\_00}, \texttt{X1\_UU}, \texttt{X1\_DD},
and \texttt{X1\_UD} -- were conducted with the \textsc{Hahndol}/\textsc{Paramesh} version
of our code using space-only adaptive mesh-refinement (AMR) with
grids placed adaptively, based on curvature invariants \cite{Baker:2006yw}.
The 4:1 mass-ratio simulation \texttt{X4\_00} was carried out with the same evolution routines, 
now ported to run within Cactus/\textsc{Carpet} \cite{cactus_web,carpet_web,etk_web}, 
which applies mesh-refinement in time as well as in space.
The initial grid structures for all runs are given in Table~\ref{table:GridData}. 
For the equal-mass simulations, the highest-resolution regions closest to the punctures had
a grid spacing of $3M/160$, $M/64$, or $3M/224$ for the medium-, high-, and ultra-high-resolution
evolutions (the ultra-high was performed for \texttt{X1\_UD} only).
For the \texttt{X4\_00} simulation, the grid resolution around the smaller hole was $M/96$,
$M/128$, and $M/160$ for the medium-, high-, and ultra-high-resolution runs.

The equal-mass simulations
exhibit between second- and fifth-order convergence for the Hamiltonian constraint, while the
momentum constraints only showed clear second-order convergence in the highest-resolution regions
around the punctures.
Nevertheless, waveform amplitudes and phases were sixth-order convergent over the majority of the
evolution. The remaining simulation, \texttt{X4\_00}, displays sixth-order convergence in
waveform amplitude and phase until close to merger time. For details, we refer the reader to
the Appendix.

We use the {\sc AHFinderDirect} code \cite{Thornburg:2003sc,Thornburg:2003sf} to locate the individual holes,
as well as the final post-merger hole. We deduce the horizon mass from the horizon area $A_{\rm AH}$ via
Christodoulou's relation \cite{Christodoulou:1970wf}
\[
\MAH^2 = \Mirr^2 + \frac{J^2}{4 \Mirr^2}\;,
\]
where $\Mirr = \sqrt{A_{\rm AH}/16\pi}$ is the \emph{irreducible mass} of the hole. We present the sum of the
two horizon masses, $\MAH \equiv M_{{\rm AH}, 1} + M_{{\rm AH}, 2}$ in Table~\ref{table:RunData}, and use it for 
time-scaling of gravitational waveforms.

Following \cite{Baker:2006kr,Husa:2007rh}, we
estimate eccentricity using the variation in puncture orbital frequency $\Omega_{\rm punc}$:
\beq
e_{\Omega}(t) \equiv \frac{(\Omega_{\rm punc}-\Omega_{\rm circ})}{2 \Omega_{\rm circ}},
\label{eq:ecc_def}
\eeq
where $\Omega_{\rm circ}$ is a monotonic fit to $\Omega_{\rm punc}$, based on a simple post-Newtonian
expansion. For a good fitting function, the residual $e_{\Omega}(t)$ should be a sinusoid of slowly
decreasing amplitude, and period equal to the orbital period; the eccentricity is then the (nearly constant)
amplitude, $e_{\Omega, \rm max}$. In practice, due in part to gauge-dependent behavior in the puncture
tracks, $e_{\Omega}(t)$ is not perfectly sinusoidal. Nevertheless, we quote the derived eccentricity
measure for each run in Table~\ref{table:RunData}. This is higher than we would like for
serious data-analysis applications, or for generating post-Newtonian--numerical-relativity hybrid waveforms, and we could choose to
reduce eccentricity through methods similar to those presented in \cite{Hannam:2010ec}. However, our
primary purpose in this paper is to investigate the bulk behavior of the waveform modes across
configurations, and very low eccentricity does not appear to be necessary for this.

To obtain gravitational waveforms from our simulations, we begin by calculating the ``outgoing radiation''
Weyl scalar $\psi_4$ \cite{Baker:2001sf}, corresponding to the tidal accelerations that are to be measured
by gravitational wave instruments. $\psi_4$ is a complex quantity related to the wave strain
$h = h_+ + i h_{\times}$ by two time-derivatives: $\psi_4 = - \ddot{h}_+ + i \ddot{h}_{\times}$ \cite{Misner73}.
We interpolate $\psi_4$ onto a set of coordinate spheres, and decompose the values on these spheres into
spherical harmonics of spin-weight $s=-2$, $\Yslm$:
\[
r \psi_4(t,r,\theta,\phi) = \sum_{\ell m} C_{\ell m}(t,r) \Yslm(\theta,\phi).
\]

To obtain the harmonic modes of the strain $h$, therefore, we must integrate $C_{\ell m}(t,r)$ twice in time,
with integration constants taken to yield zero strain long after the merger has taken place;
we call this process ``detrending'' the waveform.
Currently, we use the Fourier-domain method of time-integration presented
in \cite{Reisswig:2010di} to produce a strain waveform $h$ that is free of unwanted secular trends. This can
also be written as a sum over modes:
\beq
r h(t,r,\theta,\phi) = \sum_{\ell m} H_{\ell m}(t,r) e^{i \varphi^h_{\ell m}(t,r)} \Yslm(\theta,\phi)\;. \label{eq:strain_decomp}
\eeq

In fact, we are most interested in an intermediate quantity, $\dot{h}$, which we call
the \emph{strain-rate}. This is of particular interest because it is most closely
related to the rates of emission of gravitational-wave energy and linear momentum
\cite{Lousto:2007mh,Ruiz:2007yx}. As with the strain and $\psi_4$, the strain-rate can be decomposed
into spherical harmonics:
\beq
r \dot{h}(t,r,\theta,\phi) = \sum_{\ell m} A_{\ell m}(t,r) e^{i (\varphi_{\ell m}(t,r)+\pi/2)} \Yslm(\theta,\phi)\;, \label{eq:strainrate_decomp}
\eeq
where we have explicitly included a phase offset $\pi/2$ so that the remaining strain-rate
phase $\varphi_{\ell m}$ differs from the strain phase $\varphi^h_{\ell m}$ only by terms
of 2.5PN order (see discussion in \cite{Baker:2001sf}).

For the equal-mass cases, our extraction spheres were $\Rext \in \{45M, 50M, 55M, 60M, 65M, 70M\}$, with
a consistent extraction-region resolution of $6M/5$, $M$, and $4M/5$ for central resolutions of $3M/160$,
$M/64$, and $3M/224$, respectively. For the \texttt{X4\_00} case, the spheres were
$\Rext \in \{40M, 50M, 60M, 70M, 80M, 90M\}$. In these regions, the extraction-region resolution was $M$
and $4M/5$ for central resolutions of $M/128$ and $M/160$, respectively.

In addition to errors in the strong-field region of the source, the extracted waveforms will also contain
errors due both to finite extraction radius and finite grid resolution in the extraction region.
To mitigate the former, we have applied an extrapolation scheme to both waveform
amplitude and waveform phase. Specifically, we assume a falloff model
\beq
A_{\Rext} = A_{\infty} + \frac{a_2}{\Rext^2} \;, \; \varphi_{\Rext} = \varphi_{\infty} + \frac{f_2}{\Rext^2}.
\label{eq:Rextrap_model}
\eeq
Of all two-parameter $\Rext$-falloff models we have tried, this model gives the best fit to the amplitude
and phase of the dominant $(2,\pm2)$ modes. Adding more terms to the falloff model will introduce
overfitting errors, especially given the limited range of our $\Rext$ domain. This leading-order behavior
is consistent with $\Rext$-falloff predictions of \cite{Burko:2010au}.
This model, however, seems inappropriate for higher-frequency modes such as $(4,\pm4)$, where dissipation
effects cause a general loss in amplitude. For these, we include the possibility of an additional term
proportional to $\Rext$, at least for the amplitude:
\beq
A_{\Rext} = a_{-1} \Rext + A_{\infty} + \frac{a_2}{\Rext^2}.
\label{eq:Rextrap_model_high_m}
\eeq
We use this model for all modes with $m > 3$. We note, however, that diffusive effects should only act to
\emph{decrease} the amplitude. If a mode shows apparent \emph{growth} that does not converge with some inverse
power of $\Rext$, then it cannot be meaningfully extrapolated according to \eqref{eq:Rextrap_model_high_m}.

\section{Descriptive Radiation Characterization}
\label{sec:radiation}

As noted, the main objective of this paper is to characterize gravitational waveforms from
aligned-spin mergers. In this section we present the main features of the radiation, in a
spherical-harmonic modal decomposition. Our analysis follows the same approach developed in
\cite{Baker:2008mj}, which descriptively characterized the radiation from nonspinning
mergers in terms of an \emph{implicit rotating source} (IRS). In this approach, each modal
waveform component is viewed as the trace of the dynamic development of one of a superposed
set of source moments. To a very good approximation, each mode's radiation is circularly
polarized, indicating rotational motion. This is registered in the waveform's
\emph{rotational phase} $\Phi_{\ell m} \equiv \varphi_{\ell m}/m$, while the modal amplitudes
heuristically indicate the relative contributions of the source moments.   

Our goal is to build on the characterization of nonspinning mergers
with additional details revealing the effects of aligned spins through the merger.
We first characterize the raw content of the radiation in terms of 
energy and angular momentum, then comparatively examine how the modal amplitudes
and rotation phases develop in time.

\subsection{Radiated Energy and Angular Momentum and Final States}
\label{ssec:final_states}

To calculate the rate of energy and angular momentum emission via gravitational radiation
during merger and ringdown, we apply the following mode-summation formulas (see
Appendix~A of \cite{Baker:2008mj}):
\bea
\frac{dE}{dt}   &=& \sum_{\ell m} \frac{A_{\ell m}^2}{16\pi}\;, \label{eq:Edot}\\
\frac{dJ_z}{dt} &=& \sum_{\ell m} \frac{|m|}{16 \pi} A_{\ell m} H_{\ell m} \cos(\varphi_{\ell m}-\varphi^h_{\ell m}) \;, \label{eq:Jzdot}
\eea
where we terminate the mode-sums at $\ell=6$ for the equal-mass cases, and $\ell=5$ for
\texttt{X4\_00}. We integrate the result in time to obtain the total energy $E_{\rm rad}$
and angular momentum $J_{z, \rm rad}$ ($x$ and $y$ components are zero by symmetry) emitted during the
evolution.
In principle, these calculations could be performed with the full waveforms, rather than
the $(\ell, m)$ modes. In practice, however, we only output the mode-decomposed waveforms for
post-evolution analysis, and an accurate high-order time-integration of $\psi_4$ within the
evolution code is difficult. Additionally, post-evolution analysis with the waveform modes allows
us to better control unphysical high-frequency noise.

Rather than directly summing and integrating the $\Rext$-extrapolated strains and
strain-rates, we instead integrate the finite-$\Rext$ energy fluxes, and extrapolate the
result according to the three-parameter fit
\beq
\Delta E_{{\rm rad}, \Rext} = \Delta E_{{\rm rad}, \infty} + \frac{e_2}{\Rext^2} + \frac{e_4}{\Rext^4},
\label{eq:Erad_extrap_model}
\eeq
and similarly for the radiated angular momentum, $\Delta J_z$.
However, when dissipation effects are significant, as for the \texttt{X1\_UU} data, we must amend
this assumption according to our model \eqref{eq:Rextrap_model_high_m}. Adding an
$\Rext$-proportional term to the strain-rate amplitude will introduce several new terms to a
quadratic-in-amplitude quantity like $E_{\rm rad}$.
However, since this many terms are impossible to fit credibly with only six extraction radii,
we instead extrapolate the waveform modes first according to \eqref{eq:Rextrap_model}
 (for $m<4$) and \eqref{eq:Rextrap_model_high_m} (for $m\ge4$), and then perform a mode-sum of the result.
The results are given in Table~\ref{table:E_Jz_radiated}.

\begin{center}
\begin{table}
\caption{Radiated energy and $z$ angular momentum from all merger processes, in
units of the infinite-separation total mass estimated by $\MAH$ (final column of
Table~\ref{table:RunData}). The primary value is the $\Rext$-extrapolated value of the
$\ell=6$ mode-sum of the integrals at highest spatial resolution, while the quoted
uncertainty is the linear sum of three contributions: the standard error for the 
fit parameter for the $\ell=6$ mode-sum at the highest resolution; the difference
between the $\ell=4$ and $\ell=6$ mode-sums at this resolution; the difference between
the  $\ell=6$ mode-sum result for the highest and next-highest resolutions (for $\texttt{X4\_00}$,
the $\ell=5$ mode-sum was used instead of $\ell=6$).}
\begin{tabular}{c ll}
\hline \hline
run name        & $\Delta E_{\rm rad} (\MAH)$   & $\Delta J_{z, \rm rad} (\MAH^2)$ \\
\hline
\texttt{X1\_00} & 0.038547 $\pm$ 0.000244 & 0.367786 $\pm$ 0.001117 \\ 
\texttt{X1\_UU} & 0.075636 $\pm$ 0.001413 & 0.482665 $\pm$ 0.003874 \\ 
\texttt{X1\_DD} & 0.027240 $\pm$ 0.000219 & 0.293735 $\pm$ 0.001084 \\ 
\texttt{X1\_UD} & 0.039792 $\pm$ 0.000440 & 0.373655 $\pm$ 0.001581 \\ 
\texttt{X4\_00} & 0.014437 $\pm$ 0.000104 & 0.136347 $\pm$ 0.000853 \\ 
\hline \hline
\end{tabular}
\label{table:E_Jz_radiated}
\end{table}
\end{center}

Now we present our estimates of the final state of the post-merger Kerr holes, encoded in the
two parameters $\Mf$ and $\alpha \equiv S_z/\Mf^2$. Our estimates are derived from a number of sources, and are
tabulated in Table~\ref{table:EndStates}.

Most directly, the columns marked $M_{\rm f,rad}$ and $\alpha_{\rm rad}$ are derived from simple
conservation of energy and angular momentum:
\bea
M_{\rm f,rad}    & = & \MADM - \Delta E_{\rm rad}\;, \label{eq:Mfrad_def}\\
\alpha_{\rm rad} & = & \frac{J_0 - \Delta J_{z, \rm rad}}{M_{\rm f,rad}^2} \nonumber \\
                 & = & \frac{L_0 + S_{1z} + S_{2z} - \Delta J_{z, \rm rad}}{M_{\rm f,rad}^2}\;, \label{eq:alpharad_def}
\eea
where $\Delta E_{\rm rad}$ and $\Delta J_{z, \rm rad}$ are taken from Table~\ref{table:E_Jz_radiated},
and the remaining quantities are as in Table~\ref{table:RunData}.

We can compare with an end-state model based on fits to a range of numerical mergers.
One such model for final mass, appropriate for equal-mass systems, was given by
\cite{Reisswig:2009vc}\footnote{Other models for the post-merger mass are available;
see, for instance, Tichy \& Marronetti \cite{Tichy:2008du} and Lousto \etal
\cite{Lousto:2009mf}.}:
\beq
M_{\rm f,AEI}/\MAH = 1 - \tilde{p}_0 - \tilde{p}_1 (\alpha_1 + \alpha_2) - \tilde{p}_2 (\alpha_1 + \alpha_2)^2,
\label{eq:Mf_AEI}
\eeq
where $\alpha_A \equiv |\vec{S}_A/M_A^2|$ is the initial dimensionless spin of hole $A$,
and the fitting parameters are (again, determined by comparison with numerical data):
\bea
\tilde{p}_0 &=& 0.04826 \pm 0.00027 ,\; \tilde{p}_1 = 0.01559 \pm 0.00026 ,\nonumber \\
\tilde{p}_2 &=& 0.00485 \pm 0.00025.
\label{eq:AEI_Mcoeffs}
\eea
We note that the uncertainties on the parameters are incomplete, with an undetermined (but
presumably negligible) post-Newtonian component.

For the final spin, one model with just enough complexity for our data sets here was given by
\cite{Rezzolla:2007rz,Barausse:2009uz} \footnote{Note that we have adapted Eq.~4 of
\cite{Rezzolla:2007rz} to match our convention for $q$.}:
\bea
\alpha_{\rm AEI} &   =  & \atil + s_4 \eta \atil^2 + s_5 \eta^2 \atil + t_0 \eta \atil + 2\sqrt{3} \eta + t_2 \eta^2 + t_3 \eta^3, \nonumber\\
\atil            &\equiv& \frac{q^2 \alpha_1 + \alpha_2}{q^2+1}, \label{eq:spin_AEI} 
\eea
where the coefficients $\{s_4, s_5, t_0, t_2, t_3\}$ were determined by comparison with numerical data:
\bea
s_4 &=& -0.1229 \pm 0.0075,\; s_5 = 0.4537 \pm 0.1463,\nonumber\\
t_0 &=& -2.8904 \pm 0.0359,\; t_2 = -3.5171 \pm 0.1210,\nonumber\\
t_3 &=&  2.5763 \pm 0.4833. \label{eq:AEI_coeffs}
\eea
In Table~\ref{table:EndStates} we present final masses and spins derived from values
derived from Eqs.~\eqref{eq:Mf_AEI} and \eqref{eq:spin_AEI}, with uncertainties due only
to the parameter uncertainties in Eqs.~\eqref{eq:AEI_Mcoeffs} and \eqref{eq:AEI_coeffs}.
Unfortunately since Eq.~\eqref{eq:Mf_AEI} only applies to equal-mass cases, we cannot use it
to estimate the \texttt{X4\_00} end-state mass. The more complicated formula found in Lousto \etal
\cite{Lousto:2009mf} covers more generic binaries, but with larger uncertainties.

We note also from Table~\ref{table:EndStates} that the two cases \texttt{X1\_DD} and \texttt{X4\_00}
have the same final spin, within the quoted uncertainties. Thus we might expect
similarities in the ringdown portion of their waveforms across all important modes, though the
extent to which each quasinormal mode (QNM) is excited will be different in the two cases.

\begin{center}
\begin{table*}
\caption{End-state Kerr parameters $(M,\alpha)$ of post-merger holes. $M_{\rm f,rad}$ and
$\alpha_{\rm rad}$, and associated uncertainties, are derived from radiation balance
(\ref{eq:Mfrad_def}-\ref{eq:alpharad_def}) -- see Table~\ref{table:E_Jz_radiated}. $M_{\rm f,AH}$ and
$\alpha_{\rm AH}$ come from the \texttt{AHFinderDirect} code \cite{Thornburg:2003sc,Thornburg:2003sf}
and the \textsc{Hahndol} spin calculator \cite{coulomb}; quoted uncertainties are a combination of the post-merger
variability of the irreducible mass and spin and the difference between the measured mass and spin from
the highest and second-highest resolutions. $M_{\rm f,AEI} (\MAH)$ and $\alpha_{\rm AEI}$ use the numerically
tuned formulas \eqref{eq:spin_AEI} and \eqref{eq:Mf_AEI} due to \cite{Rezzolla:2007rd,Barausse:2009uz,Reisswig:2009vc};
quoted uncertainties here are due to uncertainty in the fitting coefficients \eqref{eq:AEI_Mcoeffs}, \eqref{eq:AEI_coeffs}.}
\begin{tabular}{c ll|ll|ll}
\hline \hline
run name        & $M_{\rm f,rad} (\MAH)$ & $\alpha_{\rm rad}$   & $M_{\rm f,AH} (\MAH)$ & $\alpha_{\rm AH}$     & $M_{\rm f,AEI} (\MAH)$  & $\alpha_{\rm AEI}$\\
\hline
\texttt{X1\_00} & 0.9519 $\pm$ 0.0002    & 0.6878 $\pm$ 0.0013  & 0.95165 $\pm$ 0.00001 & 0.68644 $\pm$ 0.00001 & 0.9517 $\pm$ 0.0003 & 0.68646 $\pm$ 0.00004\\ 
\texttt{X1\_UU} & 0.9123 $\pm$ 0.0014    & 0.9165 $\pm$ 0.0055  & 0.91164 $\pm$ 0.00013 & 0.90720 $\pm$ 0.00015 & 0.9144 $\pm$ 0.0008 & 0.9114 $\pm$ 0.0264\\ 
\texttt{X1\_DD} & 0.9645 $\pm$ 0.0002    & 0.4825 $\pm$ 0.0012  & 0.96303 $\pm$ 0.00002 & 0.48140 $\pm$ 0.00012 & 0.9637 $\pm$ 0.0006 & 0.4794 $\pm$ 0.0256\\ 
\texttt{X1\_UD} & 0.9514 $\pm$ 0.0004    & 0.6847 $\pm$ 0.0019  & 0.94996 $\pm$ 0.00001 & 0.68408 $\pm$ 0.00002 & 0.9517 $\pm$ 0.0003 & 0.6865 $\pm$ 0.0243\\ 
\texttt{X4\_00} & 0.9782 $\pm$ 0.0001    & 0.4726 $\pm$ 0.0009  & N/A                   & N/A                   & N/A                 & 0.4748 $\pm$ 0.0093\\ 
\hline \hline
\end{tabular}
\label{table:EndStates}
\end{table*}
\end{center}

\subsection{Multipolar amplitudes}
\label{ssec:amplitudes}

In Ref.~\cite{Baker:2008mj} we found strong similarity in the
peak-scaled modal amplitude development through the peak for a 
range of nonspinning mergers over a range of masses, and 
somewhat rougher similarity among the different modes.  For 
nonspinning mergers, the dominant modes were generally those
with $\ell=m$, and these modes were neatly described with
the IRS heuristic.

\begin{figure}
\includegraphics*[width=3.0in]{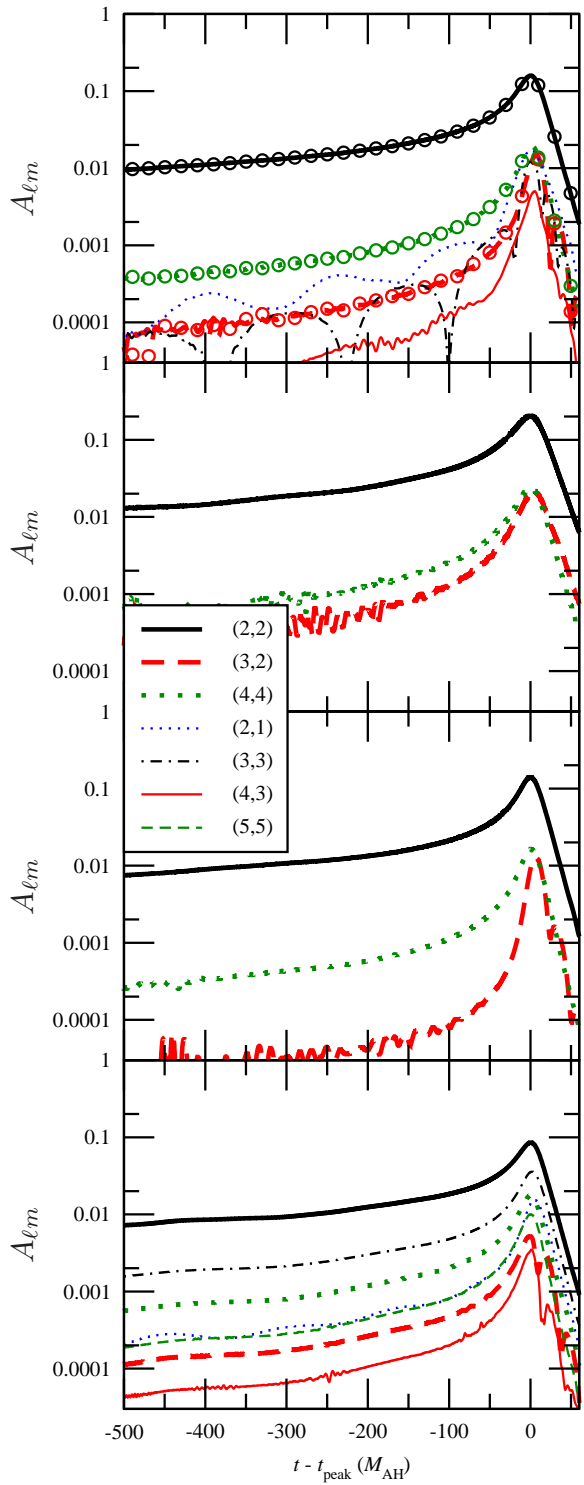}
\caption{Multipolar strain-rate amplitudes for \texttt{X1\_UD} (top panel),
\texttt{X1\_UU} (second), \texttt{X1\_DD} (third), and \texttt{X4\_00}
(bottom panel), evaluated at finite extraction radii ($45M$ for the equal-mass,
$50M$ for the \texttt{X4\_00} case). The lack of symmetry for \texttt{X1\_UD} and
\texttt{X4\_00} means additional excited (odd-$m$) modes. For the \texttt{X1\_UD}
case, we overlay circles to show the nearly identical even-$m$ mode amplitudes of
\texttt{X1\_00}. In each  panel, we omit subdominant amplitudes that never exceed
$3\%$ of the (2,2) mode.}
\label{fig:mode_amps}
\end{figure}
Strain-rate amplitudes for the strongest modes of our new simulations are shown in
Fig.~\ref{fig:mode_amps}. For all equal-mass simulations, the strongest
subdominant modes are $(4,4)$ and $(3,2)$; other modes never attain 0.1\%
of the (2,2) power (equivalently, 3\% of the (2,2) amplitude).
For \texttt{X1\_DD}, the (2,2) mode is even more
strongly dominant: in this case, all subdominant modes other than
(3,2) show significant power only at very late times.  At $R=45M$ the
(2,0) mode shows an amplitude similar to the weaker of the modes shown here,
but this is sensitive to the extraction radius (our procedure for detrending
the strain-rate doesn't work well for $m=0$).  

It is worthwhile to briefly consider how the modal composition varies with
aligned spin and mass ratio, as shown here.  Note that the (4,4) mode amplitude
is roughly the same for all cases shown here, varying even less than the (2,2)
amplitudes, as was already seen for the nonspinning runs investigated in
\cite{Baker:2008mj}. For the equal-mass 
cases, the (3,2) mode amplitude roughly equals the (4,4) mode at peak, but 
for aligned (anti-aligned) spins it is enhanced (suppressed) approaching the 
peak.  For the 4:1 mass-ratio \texttt{X4\_00}, the odd $\ell=m$ make significant contributions unseen
for equal masses, more so than in the \texttt{X1\_UD} asymmetric-spin case.  For 
asymmetric mergers of either kind, $\ell=|m|+1$ modes are also significant.

\begin{figure}
\includegraphics*[width=3.5in]{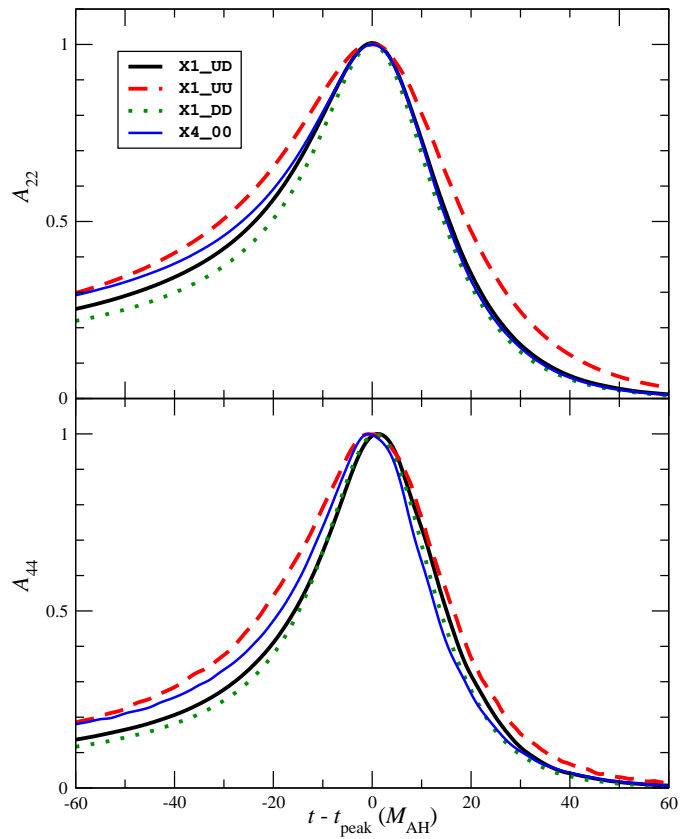}
\caption{Comparison of ``sharpness'' of amplitude peaks for $(2,2)$ (top panel)
and $(4,4)$ (bottom panel) modes of
all cases. Each amplitude has been rescaled by its maximum value.}
\label{fig:amp_22_44_ALL}
\end{figure}

Figure~\ref{fig:amp_22_44_ALL} overlays the $(2,2)$ (top panel) and $(4,4)$
(bottom panel) amplitude peaks of all cases (suitably time-shifted and rescaled)
to compare their relative sharpness.
It is interesting that the ``down-down'' peaks of \texttt{X1\_DD} are narrower
than the ``up-up peaks'' of \texttt{X1\_UU}.  The steeper slope on the
$t>0$ side can be tied to the generally faster fall-off in QNM modes for the 
prograde modes of the much more slowly rotating black hole generated by the 
down-down merger. The peaks remain roughly symmetric, with a faster rise as
well.  This is particularly striking for the subdominant modes. 

\subsection{Waveform phasing}
\label{ssec:phasing}

In our studies of nonspinning mergers \cite{Baker:2008mj}, we found strong
correspondence in phase development among the different modes, interpreted as
near-``corotation'' of the implicit-source moments. Specifically, all significant
modes displayed a common rotational phase $\Phi_{\ell m}$ up to the time of peak
power at merger, deviating by less than 0.025 rad during that time. After the
merger, the deviations between modes increased, but for the $\ell=m$ modes this
deviation was very slow, $\lesssim$ 1 rad over the first $100M$ following merger.
For the weaker $\ell\neq m$ modes, the phasing began to differ somewhat earlier and was
in some cases less cleanly described by the IRS heuristic.

\begin{figure*}
\includegraphics*[width=3.0in]{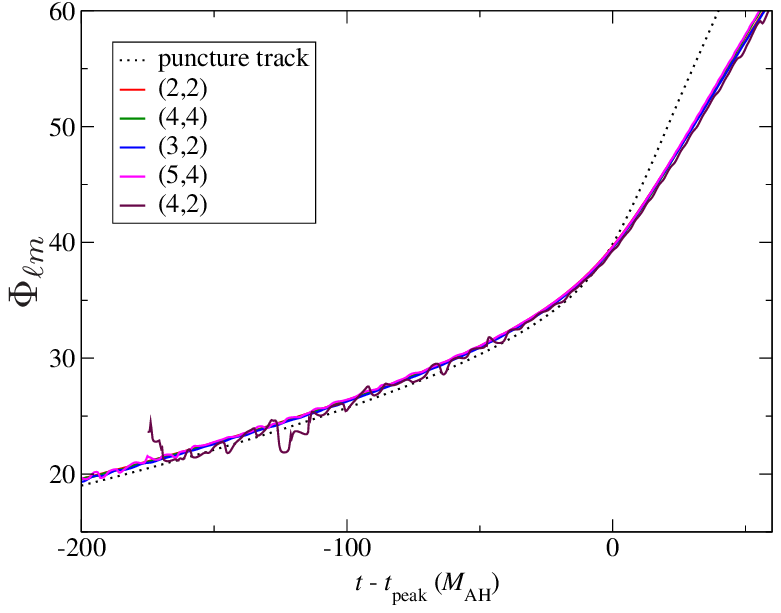}
\includegraphics*[width=3.0in]{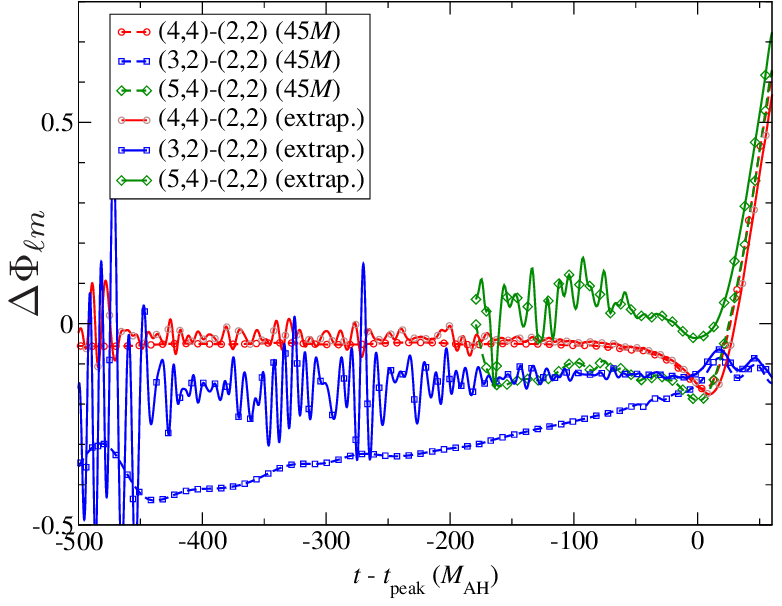}
\caption{Left: Rotational phase $\Phi(t)$ from puncture tracks and different multipolar
strain-rate components extrapolated to $\Rext \rightarrow \infty$ for the
up-up case \texttt{X1\_UU}. The weakest two modes, $(5,4)$ and $(4,2)$, are only
measurable for about $200 M$ before merger.
Right: The difference in phase with the $(2,2)$ mode for the next three most important
modes only: $(4,4)$ (circles), $(3,2)$ (squares), and $(5,4)$ (diamonds). In each case, we
show the difference at $\Rext=45M$ (dashed lines), and $\Rext \rightarrow \infty$ (solid lines).}
\label{fig:phase_UU}
\end{figure*}
Modal phase comparisons are more challenging for equal-mass spinning mergers
than they were for nonspinning unequal-mass mergers.  As noted above, the symmetries of
the configuration and the weakness of radiative spin effects in the inspiral
yield only a few significant modes and even these tend to be weak, subject to
competition with noise in the simulations, and likely more sensitive to
subtleties in the choice of spherical-harmonic basis.

The left panel of Fig.~\ref{fig:phase_UU} shows the phasing of several strain-rate
modes for the up-up case \texttt{X1\_UU}.
Generally, as was the case for nonspinning systems \cite{Baker:2008mj}, the
different $(\ell,m)$ modes remain approximately in phase up to the merger. As
with the nonspinning mergers, the $\ell=m$ cases show the best agreement for
$t<0$. In this case though, because exchange symmetry excludes the odd-$m$
modes, we only have two $\ell=m$ modes to compare up to $\ell=5$. Two modes
present -- $(4,2)$ and $(5,4)$ -- have amplitudes below our 3\% cut-off in
Fig.~\ref{fig:mode_amps}. Such small amplitudes introduce a lot of noise in the
mode's phase; we include the phase only when it begins to show acceptable
continuity.

We take a closer look at the relative phasing in the right panel of
Fig.~\ref{fig:phase_UU}, where we present the difference between each of the three
strongest subdominant modes --  $(4,4)$, $(3,2)$, and $(5,4)$ -- with the dominant
$(2,2)$ mode. Generally the phase differences decrease going from a finite extraction
radius (dashed curves) to $\Rext \rightarrow \infty$ (solid curves).

Looking at the inspiral portion ($t<0$) of the phase first, we see that the $(2,2)$
and $(4,4)$ rotational phases agree within $\sim 0.05$ rad, with a marginal
improvement when we extrapolate $\Rext \rightarrow \infty$. For the $(3,2)$ mode,
there is a roughly constant offset of about $0.15$ rad after $\Rext$-extrapolation.
Due to the short extent and noisy nature of the reliable $(5,4)$ mode phase, it is
difficult to extract a definite phase offset; it appears to be in the range
$\sim 0.05 - 0.10$ rad after extrapolation. 
However, the phase offset has also flipped
sign during extrapolation in $\Rext$, indicating that we may not know the correct
phase to high accuracy.
This may not be resolved simply by increasing grid resolution; we have seen similar
extrapolation sign-flips for the $(5,4)$ phase offset in our \texttt{X1\_UD} simulations,
even at the ``ultra-high'' $3M/224$ resolution.
As with the nonspinning case, the $\ell\neq m$ modes show the largest offset from the
$(2,2)$ modes, and are most affected by $\Rext$-extrapolation effects.

We note that post-Newtonian theory predicts for nearly constant phase offsets between
modes during late inspiral; these come in at 1.5PN order for certain modes (see, for
example, the polarization amplitudes given in \cite{Kidder:2007rt}). However, they are
small compared to the phase differences shown here -- less than $\sim 0.03$
rad up to $100M$ before peak.

Looking now at the post-merger period ($t>0$), the phase agreement remains quite
tight, better than that seen in the nonspinning mergers. In the IRS interpretation,
all modes in this case remain nearly rotationally locked right through the merger.
Note in particular that the phase difference between the $(2,2)$ and $(3,2)$ modes is
roughly constant for $t>0$.
The $(4,4)$ and $(5,4)$ modes are also in phase with each other at $\Rext=45M$; they
develop a phase offset when extrapolated to $\Rext\rightarrow\infty$, but maintain the
same slope. All modes agree within $\lesssim 1$ rad even $60M$ after peak. 

To understand this tight phase agreement, we may look to perturbation theory for the
post-merger Kerr hole. For rapidly spinning black holes, QNM frequencies depend
primarily on $m$, approaching $M\omQ=m/2$ in the $a\rightarrow \Mf$
limit \cite{Detweiler:1980gk,Berti:2009kk}, which suggests a tighter coupling for the
modes in this case. However, the final spin of the post-merger Kerr hole for
\texttt{X1\_UU}, $\alpha \approx 0.91$, is not close enough to this extremal limit to
explain the phase agreement we see.

\begin{figure*}
\includegraphics*[width=3.0in]{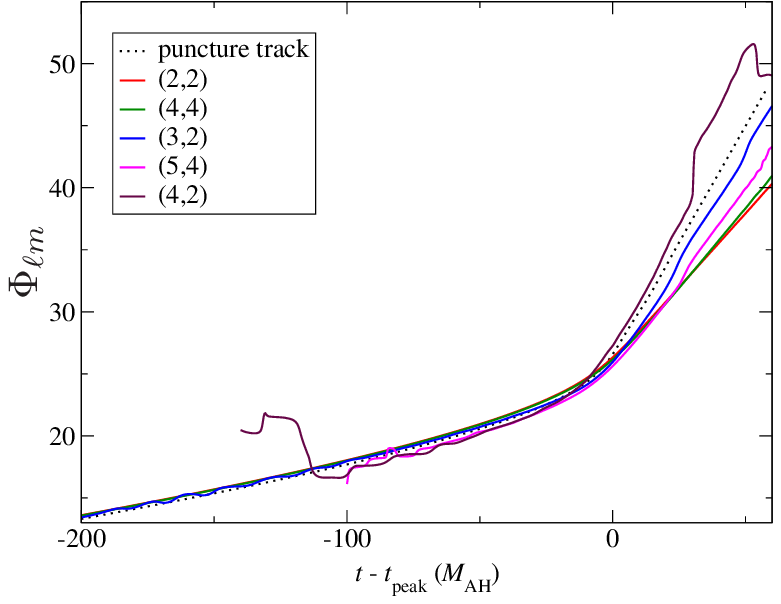}
\includegraphics*[width=3.0in]{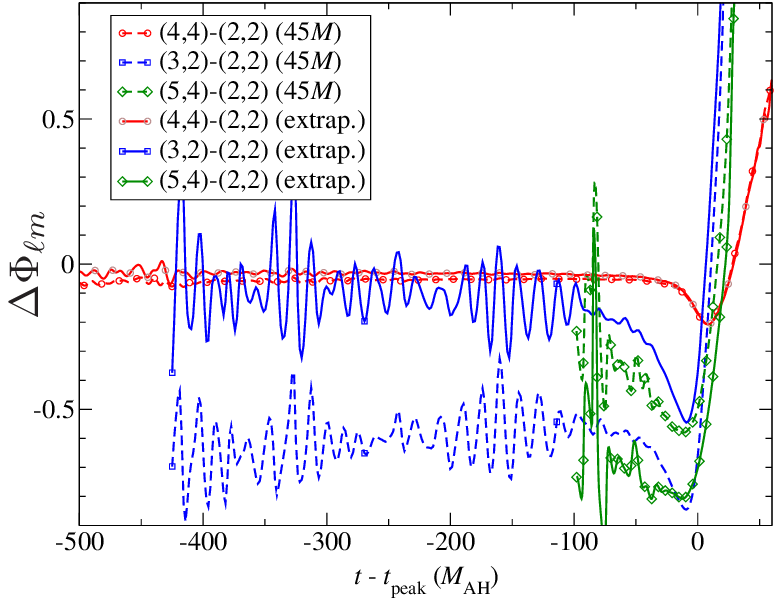}
\caption{Left: Rotational phase $\Phi(t)$ from puncture tracks and different multipolar
strain-rate components extrapolated to $\Rext \rightarrow \infty$ for the down-down case
\texttt{X1\_DD}. The weakest two modes, $(5,4)$ and $(4,2)$, arise very suddenly at late
times, and are only measurable for about $100 M$ before merger.
Right: The difference in phase with the $(2,2)$ mode for the next three most important
modes only: $(4,4)$ (circles), $(3,2)$ (squares), and $(5,4)$ (diamonds). In each case, we
show the difference at $\Rext=45M$ (dashed lines), and $\Rext \rightarrow \infty$ (solid lines).}
\label{fig:phase_DD}
\end{figure*}

In the left panel of Fig.~\ref{fig:phase_DD} we show the phasing of several modes for the
down-down simulation \texttt{X1\_DD}. Again, the $(4,2)$ and $(5,4)$ modes are weak, and
yield reliable phases only from $\sim 100M$ before peak. Looking at the right panel, the
$(4,4)$ mode extrapolated phase difference is $\sim 0.02$ rad, while there is a small offset
of about $0.1$ rad between the $(3,2)$ and $(2,2)$ modes. As with the \texttt{X1\_UU} mode,
extrapolation in $\Rext$ appears to increase the phase offset.
There is a slight drift among the $\ell=m$ modes, similar to that
seen in other cases. Indeed, the phase difference between (2,2) and (4,4) is
nearly identical to that of the \texttt{X1\_UU} case. 
The $\ell\neq m$ modes this time show significantly varying frequencies
(slopes). This is unsurprising, consistent with the differences among the leading
normal QNM mode frequencies.

\section{Modeling the Near-Merger Waveform}
\label{sec:model}

In this section we undertake a more quantitative study of the late-time
waveforms through merger and ringdown, following the general approach in 
\cite{Baker:2008mj} for an explicit quantitative representation of the frequency
development.  We extend the previous work on modeling the amplitude with additional
parameters to allow more precise fits at relatively early times.

The waveform phasing examined in the last section is fairly featureless. The 
phase is monotonic, slowly developing curves with a gentle elbow at merger.  
This simplicity is a result of the slow secular development of the underlying 
circular motion which generates the radiation.  It also suggests that we may
quantify the phase development with just a few parameters.  

Following the approach in \cite{Baker:2008mj} we probe more deeply into the phasing by
taking a time-derivative to study the frequency evolution.  Common features
are found among the leading waveform modes and across a range of mergers, allowing
the results to be summarized with a simple parametrization.
With the same general frequency model as in \cite{Baker:2008mj} for nonspinning mergers
we can also describe the phasing of spinning black-hole mergers.

\begin{figure}
\includegraphics*[width=3.5in]{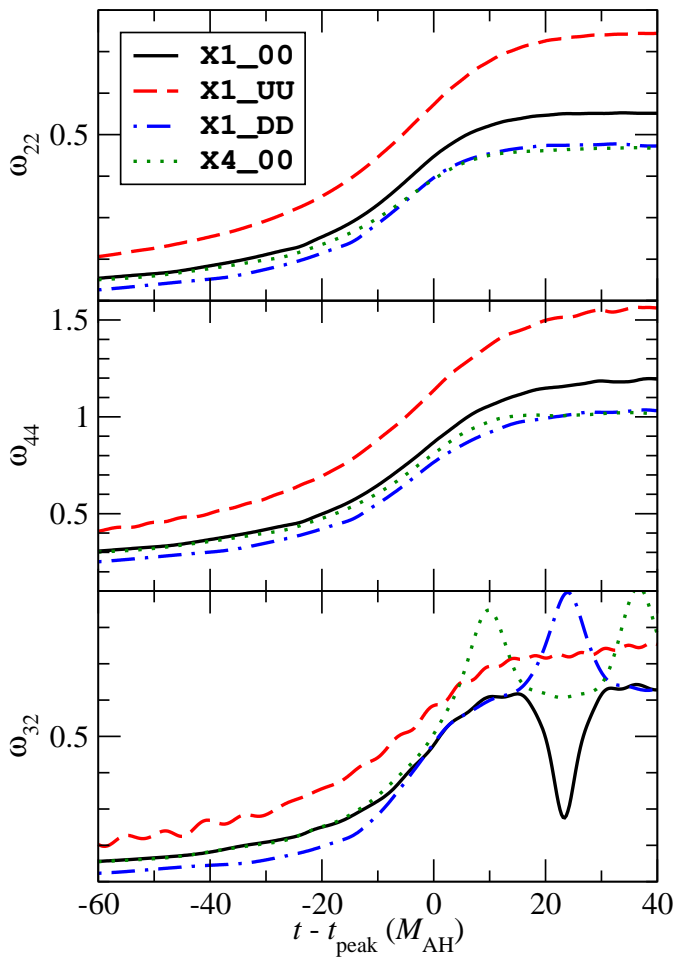}
\caption{Waveform frequency $\omega_{\ell m}$ for the dominant modes of the equal-mass
evolutions. The upper panel shows the dominant $(2,2)$ mode, while the middle and bottom
panels show the next strongest modes -- $(4,4)$ and $(3,2)$, respectively. We also show
the corresponding frequencies for the 4:1 nonspinning merger \texttt{X4\_00}. At early
times, this tracks the \texttt{X1\_00} waveform, while during merger it approaches the final
frequency of the \texttt{X1\_DD} case.} 
\label{fig:omega_comps}
\end{figure}

In Fig.~\ref{fig:omega_comps}, we compare the dominant-mode frequencies of the three
equal-mass cases presented above. Since odd-$m$ modes are suppressed by symmetry, the
relevant modes are the $(2,2)$ (quadrupole), $(4,4)$, and $(3,2)$ modes. Unsurprisingly, the
frequencies are consistently higher throughout merger for more-aligned spins, with
the final plateau value matching the dominant quasinormal-mode (QNM) frequency. For the 
$(3,2)$ modes, there is significant deviation from the smooth frequency development
generally expected according to our IRS heuristic; this amounts to a large bump
in the frequency during the plateau phase.  Similar deviations in the $(3,2)$ modes have
been noted previously \cite{Buonanno:2006ui,Baker:2008mj}.  Such effects may arise through mode 
mixing with the $(2,2)$ mode \cite{Pan:2011gk}, which could arise through ambiguity in the
shape of the sphere on which the radiation is measured, or on the use of (spin-weighted)
spherical harmonics, rather than the spheroidal harmonics appropriate for the perturbation
theory in which the QNM frequencies are defined \cite{Teukolsky:1973ha}. The precise cause
and mechanisms of this mixing are open questions, which we hope to return to in future work.

Also included in Fig.~\ref{fig:omega_comps} are the equivalent frequencies for the
4:1 nonspinning merger \texttt{X4\_00} (note, however, that that merger had significant
odd-$m$ modes not
present in the equal-mass cases here). The spins of the anti-aligned \texttt{X1\_DD}
initial data were chosen to yield the same final Kerr parameters (mass, spin) as the
\texttt{X4\_00} data, according to \eqref{eq:spin_AEI}. As the Kerr parameters
determine the QNM frequencies of each mode, it is not surprising that the \texttt{X1\_DD}
and \texttt{X4\_00} frequencies level off to the same value after merger. What is
interesting is the difference in behavior approaching this final state. For
$t \lesssim -20 M$, \texttt{X4\_00} hews closely to the nonspinning \texttt{X1\_00}.
The latter could be expected given the similarity in phasing upon approach to merger for
nonspinning mergers \cite{Baker:2008mj}. At the latest times the frequency development
is determined primarily by the parameters of the final black hole formed, while additional
parameters become important as we look back to earlier times.

\subsection{Modeling the Rotational Frequency}
\label{ssec:freq_model}

In \cite{Baker:2008mj}, we introduced the following empirical model for the rotational frequency
$\Omega_{\ell m} \equiv \omega_{\ell m}/m$ in a short time-window around the merger:
\bea
\Omega(t)&=& \Of(1-\hat f(t))\label{eq:OmegaModel1}\\
\hat f(\kappa,b,t_0;t)&=& \frac{c}{2} \left(1 + \frac{1}{\kappa}\right)^{(1+\kappa)} \nonumber \\
                      & & \times \left[1 - \left( 1 + \frac{1}{\kappa} e^{-2(t-t_0)/b} \right)^{-\kappa}\right],
\label{eq:OmegaModel}
\eea
where the dimensionless parameter $c=\Odotmax b/\Of$ replaces the ``maximum frequency slope''
$\dot{\Omega}_0$ corresponding to the peak chirp rate.
Within this general framework, there are up to five free parameters for the frequency development: $c$,
$\kappa$, $b$, $t_0$, and $\Of$.

In our previous investigations \cite{Baker:2008mj}, this functional form worked well in fitting
the dominant frequencies of a sequence of nonspinning binaries with mass ratios in the range
$\{1.0,6.0\}$. Unsurprisingly, $\Of$
was found to be consistent with the quasinormal frequency of the post-merger Kerr hole. 
More interestingly, $b$ was also found to be approximately consistent with the 
quasinormal damping time, meaning that at late times the frequency approaches its limiting value 
exponentially at the same rate as amplitude squared.
It was also found that the dimensionless ratio $\Mf \Odotmax/\Of \approx 0.021$ across all cases.

We usually apply \eqref{eq:OmegaModel1} as an ``orbital frequency'', which is scaled from the gravitational-wave
frequency by the azimuthal mode number $m$. The formula may be applied, with similar results, to
strain, strain-rate or $\psi_4$ waveforms.

We consider three increasingly constrained classes of fits of this form. 
The most general is a free fit for all five parameters.
Second, we test the late-time frequency/amplitude relationship noted in \cite{Baker:2008mj} with a fit
where  $b$ is constrained  to agree with the late-time amplitude fall-off rate (and thus with the QNM 
fall-off rate). Finally we suggest a general fit by which all parameters (other than $t_0$) are 
derived from the final black-hole mass and spin.

\begin{figure}
\includegraphics*[width=3.5in]{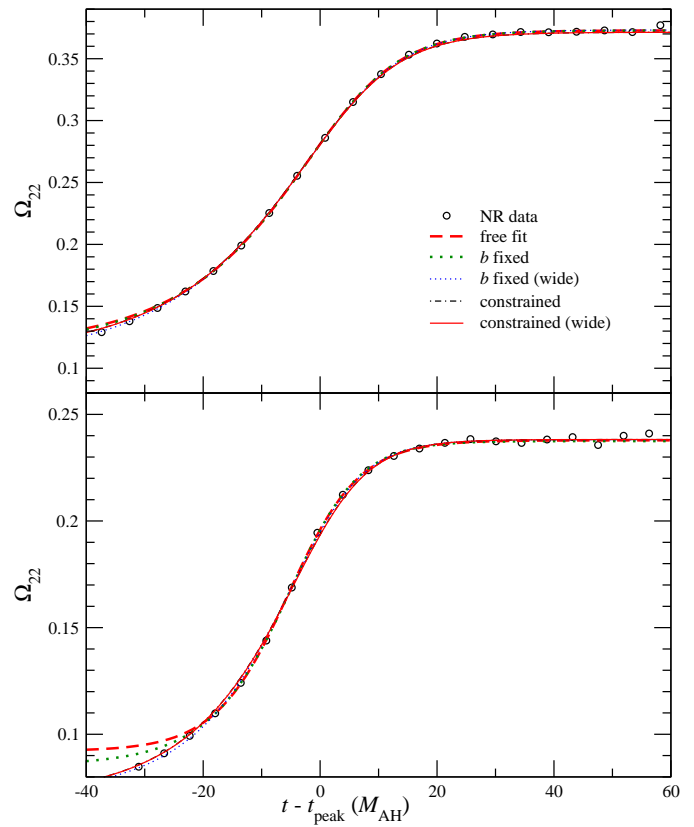}
\caption{Rotational frequency fits for the $(2,2)$ modes of the \texttt{X1\_UU} (top) and
\texttt{X1\_DD} (bottom) runs, extrapolated to $\Rext \rightarrow \infty$.
The different fits differ primarily in the early part of the comparison window around
$t_{\rm peak}$, which runs from $- 20 \MAH$ to $+ 40 \MAH$ for standard fits, and from
$- 40 \MAH$ to $+ 60 \MAH$ for wide fits.}
\label{fig:freq_fits_UU_DD}
\end{figure}

We focus on the strain-rate fit, as it provides a good compromise between the base-line
drift error that affects the strain waveforms, and the higher level of noise in the
$\psi_4$ waveforms. Figure~\ref{fig:freq_fits_UU_DD} shows the result of this procedure for
the $\Rext$-extrapolated waveforms of the \texttt{X1\_UU} (top) and \texttt{X1\_DD} (bottom)
runs (note that the numerical data have been down-sampled by factors of 20 or more for clarity of
presentation). At the level of precision accessible by eye,
all fits appear nearly perfect after $t>-20$. Parameter fits conducted only over times
$t>-20$ typically do not extrapolate well to earlier times; a fit over a wider range, extending
over $t>-40$ appears to be very good over this entire region though there is some slight degradation 
in the quality of the fit near $0<t<10$. For the dashed curves labeled ``free fit'', all five
parameters in \eqref{eq:OmegaModel} were fit freely against the numerical data. In the curves labeled
``$b$ fixed'', we test the hypothesis that the exponential decay of frequency evolution is related to
the amplitude fall-off rate, fixing $b$ first by a fit to the exponential decay rate in the 
mode amplitude data, before fitting the other parameters according to the frequency data. Though the
frequency data are then fit against four parameters instead of five, the result is still a
very good fit, justifying the assumption. The amplitude/frequency relationship is discussed more 
in the next section.
With $b$ based on the amplitude data, we record the best-fit values of the parameters $c$, $\kappa$,
$b$, $\Of$, and $t_0$ in Table~\ref{table:fit_params_free}. The final curves in
Fig.~\ref{fig:freq_fits_UU_DD} (labeled ``constrained'') only fit $t_0$, with all other parameters
pre-set, as discussed below in Sec.~\ref{ssec:constrained_model}.

\subsection{Amplitude Modeling}
\label{ssec:amp_model}

Following \cite{Baker:2008mj}, our strategy is to describe the wave amplitudes in relation to the
frequency. This is loosely motivated by the idea that frequency evolves in response to loss of
energy and angular momentum, but energy and angular momentum fluxes are dependent on the wave
amplitude. For nonspinning systems, we previously found that $dJ/d\Omega$ was slowly varying
in the merger, and could be approximated as a constant.
The result was enough to provide a coarse quantitative description of the merger-ringdown
amplitude development in reference to the frequency development.  Here we extend that
model, introducing additional parameters to allow more precise quantitative description 
of the numerical results.

Now assume the waveform strain-rate amplitude takes the general form:
\beq
A_{\ell m}(\hat f) = A_{0 \; \ell m} \,\afunc(\hat f) \sqrt{\left|\dot{\hat f}(t)\right|},
\label{eq:AmpModel_strainrate}
\eeq
where the adjusting function $\afunc(x)$ is some function that approaches
unity as $x\rightarrow0$.
Then assuming that the strain amplitude $H_{\ell m}$ satisfies
$|\dot{H}_{\ell m}| \ll |H_{\ell m} \dot{\varphi}^h_{\ell m}|$, which is true at all points of the
numerical waveform, we can find an approximate expression linking strain and strain-rate:
\beq
A_{\ell m} \approx H_{\ell m} \omega^h_{\ell m}.
\eeq
Combining this with \eqref{eq:AmpModel_strainrate}, we can model the \emph{strain}
amplitude as:
\bea
H_{\ell m}(\hat f) &=& \omega_{\ell m}^{-1} \; A_{\ell m}(\hat f) \nonumber \\
                   &=& (m \Of)^{-1} (1-\hat f)^{-1} \; A_{0 \, \ell m} \; \afunc(\hat f) \sqrt{\left|\dot{\hat f}(t)\right|} \nonumber \\
                   &=& H_{0 \, \ell m} \; (1-\hat f)^{-1} \afunc(\hat f) \sqrt{\left|\dot{\hat f}(t)\right|}. 
\label{eq:AmpModel_strain}
\eea
Similarly, the amplitude of the $\psi_4$ $(\ell,m)$ mode would be
modeled as:
\beq
C_{\ell m}(\hat f) = C_{0 \, \ell m} \; (1-\hat f) \afunc(\hat f) \sqrt{\left|\dot{\hat f}(t)\right|}, 
\label{eq:AmpModel_psi4}
\eeq
where we are still using the strain sign convention for phasing of modes (that is,
positive-$m$ modes have positive frequencies). 

Parameters including $A_0$ and other parameters in the definition of $\afunc(\hat f)$ allow
some tuning for various cases considered here while preserving the general approach in
\cite{Baker:2008mj}.
Concretely, consider
\beq
\afunc(\hat f)^{-2}=1+\sum_{n=1}^N \alpha_n\left(\hat f^{2n}-\hat f^{2n+2}\right).
\label{eq:Qfac_quad}
\eeq
The simplest possibility, with $N=0$ yielding $\afunc=1$, would imply that $dE/d\omega$
is constant, i.e. that the system loses radiative energy in linear proportion to the
late-time frequency decay to the quasinormal-ringdown rate; this is close to the amplitude
model used in \cite{Baker:2008mj}.
Equation~\eqref{eq:Qfac_quad} is consistent with
quasinormal ringing radiation in the $f\rightarrow0$ limit (assuming frequency model
parameter $b=1/\Im{\omQ}$) and can be adjusted for deviations earlier in the waveform
where $1>f>0$. The restriction to even powers was motivated by an empirical observation
that the first helpful correction seems to be at second order, and the resummed powers
in the summand, yields a more generally regular result as $f\rightarrow1$. Going back to
times more than $20M$ before merger the model becomes unrealistic. The model amplitude
begins small at early times, growing exponentially toward the peak.

\begin{figure}
\includegraphics*[width=3.5in]{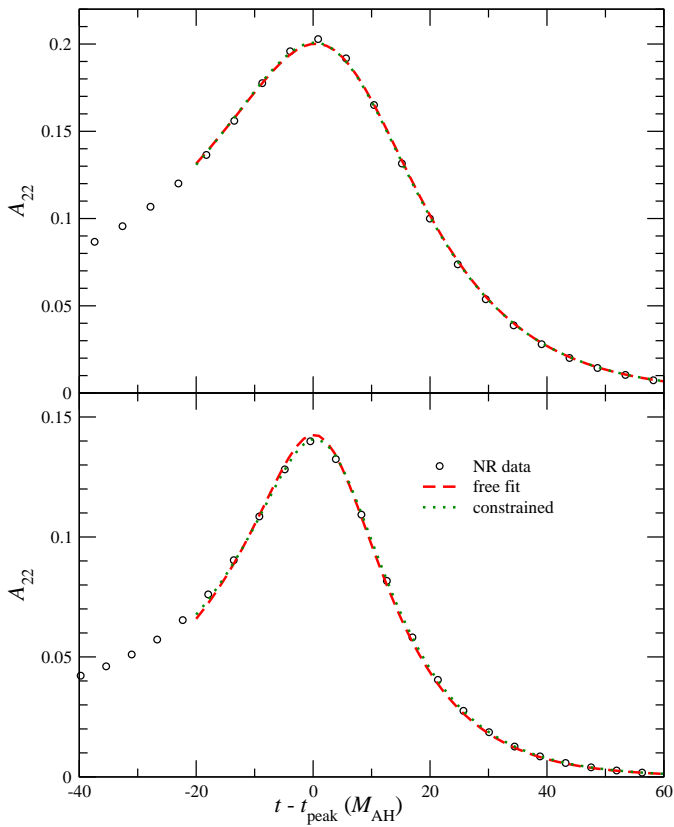}
\caption{Amplitude fits for the $(2,2)$ modes (extrapolated to $\Rext \rightarrow \infty$) of
the \texttt{X1\_UU} (top) and \texttt{X1\_DD} (bottom) runs.
The two fits in each panel differ in whether they fit the parameter $\alpha_1$ or just $A_0$.
The fit windows here emphasize the late tail of the amplitude, running from $- 20 \MAH$ to $+ 120 \MAH$.}
\label{fig:amp_fits_UU_DD}
\end{figure}

In practice, we find that we get a good approximation for the merger-ringdown part of
the radiation in the (2,2) modes by keeping one term in the expansion \eqref{eq:Qfac_quad},
and fitting for $A_0$ and $\alpha_1$. Then using $\hat f$ from \eqref{eq:OmegaModel},
the complete amplitude model used is
\beq
A_{22}^2 \equiv |r \dot h_{22}(\hat f)|^2 = A_0^2 \,\frac{\dot{\hat f}(t)}{1+\alpha_1\left(\hat f^{2}-\hat f^{4}\right)}.
\label{eq:AmpModel22_strainrate_trunc}
\eeq
The result of this procedure is shown in Fig.~\ref{fig:amp_fits_UU_DD},
for the  $\Rext$-extrapolated waveforms of the \texttt{X1\_UU} and \texttt{X1\_DD} runs.
For the less-constrained fits, $b$ was first determined using data in a window
from $20 \MAH$ to $80 \MAH$; this value was then fixed, and data over the larger
window $-20\MAH$ to $110\MAH$ were used to determine $A_0$ and $\alpha_1$.
Though not shown in the figure, we see somewhat less accurate fits with
the model for the amplitudes of the \texttt{X4\_00} case, with differences
before and near peak at the $\sim 5\%$ level.

\subsection{Constraining the Models}
\label{ssec:constrained_model}

Working with the results of the free fits for frequency and amplitude, we note
the approximate constancy of the parameters $c$ and $\kappa$ across all cases.
Additionally, we note that the final frequency and decay parameters $\Of$ and
$b$ are close to the expected QNM values. Thus
we may be able to reduce considerably the number of free parameters needed for
the models.

Similar to the dimensionless scaling used for the $c$ parameter above we seek
a scaling of $\alpha_1$ in terms of the QNM ``quality factor''
$Q=\Re\omQ/2 \Im\omQ$, as a simple dimensionless number dependent on the spin
of the final hole.
We find that the results for our equal-mass cases roughly scale with $Q^2$, with
the mean result $\alpha_1\sim72.3/Q^2$.

We now perform a more constrained version of the fits, fixing the parameters $b$,
$\Of$ to their QNM values, and replacing $c$ and $\kappa$ with their average values
from Table~\ref{table:fit_params_free}, and setting the frequency parameters and
$\alpha_1$ as outlined above. The combined set of constrained parameters is:
\bea
c &=& 0.252 \; , \kappa = 0.426 \; , \Of = \Re\omQ/2 \; , \nonumber \\
b &=& 1.0/\Im\omQ \; , \alpha_1 = 72.3/Q^2.
\label{eq:IRS_params_const_22}
\eea
Thus we are left with just two free parameters to fit: $t_0$ and $A_0$. These
(as well as the constrained parameters) are recorded in
Table~\ref{table:fit_params_constrained}. 

We may consider attempting to constrain the remaining parameters as well.
The $A_0$ parameter has units $\Mf^{1/2}$ and seems to scale approximately with
$\Of^{-1/2}$. Using the mean fit for the equal-mass cases we get
\beq
A_0\sim9.9\eta\Of^{1/2}.
\label{eq:IRS_A0_const_22}
\eeq
We include $\eta$ in the fit since the overall amplitude
coefficient must vanish linearly as the mass-ratio goes to zero. Though we have not
focused on mass-ratio dependence, this scaling is consistent with the nonspinning
4:1 result.

\begin{center}
\begin{table*}
\caption{Best-fit values for the unconstrained parameters $c$, $\kappa$, $b$, $\Of$, and
$t_0$ for the frequency model \eqref{eq:OmegaModel1}, and of $A_0$ and $\alpha_1$ for the
amplitude model \eqref{eq:AmpModel_strainrate}. All fits are over a time window from
$t_{\rm peak} - 40 M$ to $t_{\rm peak} + 60 M$. Quoted uncertainties are the direct sum
of three terms: uncertainties in the highest-resolution fits; differences between best-fit
values for $\Rext \rightarrow \infty$ and $\Rext = 45M$ ($40M$ for \texttt{X4\_00}); differences
between best-fit values at highest and next-highest resolutions runs.}
\begin{tabular}{c lllllll}
\hline \hline
run name        & $c$                 & $\kappa$          & $t_0$            & $b$                & $\Of$                 & $A_0$             & $\alpha_1$ \\
\hline
\texttt{X1\_00} & 0.2489 $\pm$ 0.0040 & 0.421 $\pm$ 0.015 & -3.77 $\pm$ 0.17 & 11.685 $\pm$ 0.025 & 0.27655 $\pm$ 0.00021 & 1.270 $\pm$ 0.022 &  6.64 $\pm$ 0.35\\ 
\texttt{X1\_UU} & 0.2500 $\pm$ 0.0021 & 0.401 $\pm$ 0.009 & -2.15 $\pm$ 0.59 & 14.296 $\pm$ 0.030 & 0.37317 $\pm$ 0.00070 & 1.627 $\pm$ 0.064 &  2.46 $\pm$ 0.70\\ 
\texttt{X1\_DD} & 0.2626 $\pm$ 0.0062 & 0.473 $\pm$ 0.027 & -4.90 $\pm$ 0.12 & 11.203 $\pm$ 0.034 & 0.23805 $\pm$ 0.00093 & 1.157 $\pm$ 0.009 & 11.12 $\pm$ 0.12\\ 
\texttt{X1\_UD} & 0.2458 $\pm$ 0.0038 & 0.407 $\pm$ 0.014 & -3.57 $\pm$ 0.15 & 11.634 $\pm$ 0.024 & 0.27643 $\pm$ 0.00017 & 1.263 $\pm$ 0.016 &  6.27 $\pm$ 0.38\\ 
\texttt{X4\_00} & 0.2343 $\pm$ 0.0033 & 0.439 $\pm$ 0.009 & -4.73 $\pm$ 0.52 & 11.381 $\pm$ 0.033 & 0.23380 $\pm$ 0.00075 & 0.740 $\pm$ 0.017 &  9.96 $\pm$ 1.26\\
\hline \hline
\end{tabular}
\label{table:fit_params_free}
\end{table*}
\end{center}

\begin{center}
\begin{table*}
\caption{Values for the parameters $c$, $\kappa$, $b$, $\Of$, and $t_0$ for the frequency
model \eqref{eq:OmegaModel1}, and of $A_0$ and $\alpha_1$ for the amplitude model
\eqref{eq:AmpModel_strainrate}. Unlike in Table~\ref{table:fit_params_free}, only $t_0$
and $A_0$ are freely fit; the remaining parameters have been fixed, as given in
Eq.~\eqref{eq:IRS_params_const_22}. All fits are over a time window from
$t_{\rm peak} - 40 M$ to $t_{\rm peak} + 60 M$. Quoted uncertainties are the direct sum
of three terms: uncertainties in the highest-resolution fits; differences between best-fit
values for $\Rext \rightarrow \infty$ and $\Rext = 45M$ ($40M$ for \texttt{X4\_00}); differences
between best-fit values at highest and next-highest resolutions runs.}
\begin{tabular}{c lllllll}
\hline \hline
run name        & $c$   & $\kappa$ & $t_0$            & $b$    & $\Of$   & $A_0$               & $\alpha_1$ \\
\hline
\texttt{X1\_00} & 0.252 & 0.426    & -3.99 $\pm$ 0.29 & 11.712 & 0.27661 & 1.271 $\pm$ 0.013 &  6.7934 \\ 
\texttt{X1\_UU} & 0.252 & 0.426    & -2.31 $\pm$ 0.58 & 14.404 & 0.37133 & 1.633 $\pm$ 0.032 &  2.4924 \\ 
\texttt{X1\_DD} & 0.252 & 0.426    & -4.41 $\pm$ 0.36 & 11.222 & 0.23820 & 1.143 $\pm$ 0.006 &  9.9784 \\ 
\texttt{X1\_UD} & 0.252 & 0.426    & -3.91 $\pm$ 0.30 & 11.681 & 0.27661 & 1.272 $\pm$ 0.006 &  6.8296 \\ 
\texttt{X4\_00} & 0.252 & 0.426    & -6.67 $\pm$ 0.72 & 11.465 & 0.23305 & 0.732 $\pm$ 0.009 &  9.9867 \\
\hline \hline
\end{tabular}
\label{table:fit_params_constrained}
\end{table*}
\end{center}

\subsection{Subdominant modes}

\begin{figure}
\includegraphics*[width=3.5in]{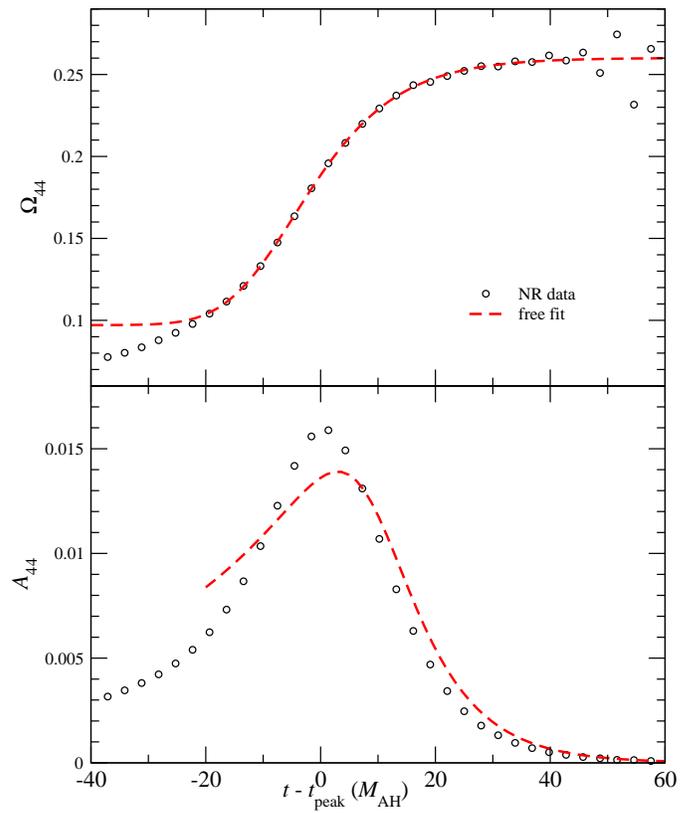}
\caption{Frequency (top panel) and amplitude (bottom panel) fits for the $(4,4)$ mode
(extrapolated to $\Rext \rightarrow \infty$) of the \texttt{X1\_DD} run. Again, the
numerical data are indicated by circles, with a free fit over the parameters represented by the
dashed line.}
\label{fig:freq_amp_fits_DD_44}
\end{figure}

As noted above the most significant modes for equal-mass mergers are the $(4,4)$ and $(3,2)$ modes.
Even these have amplitudes of no more than about one-tenth that of the $(2,2)$ mode. While the
$(3,2)$ mode shows more complicated features that do not lend themselves to this fitting analysis, the 
$(4,4)$ mode is phenomenologically similar to the $(2,2)$ mode. 
Figure~\ref{fig:freq_amp_fits_DD_44} shows the frequency and amplitude fits for the $(4,4)$ mode
of the \texttt{X1\_DD} run. The frequency fit is clearly still very close to the numerical data
over the domain of interest, but the amplitude's overall peak is $\sim$ 10\% too low, with a poor
fit to the slope of the numerical data before the peak. This suggests that our ansatz for the
mode amplitude does not carry over to subdominant modes, and requires further work. Nevertheless, we
will see in the next section that the dominant mode may already be useful in detection studies.

\begin{figure}
\includegraphics*[width=3.5in]{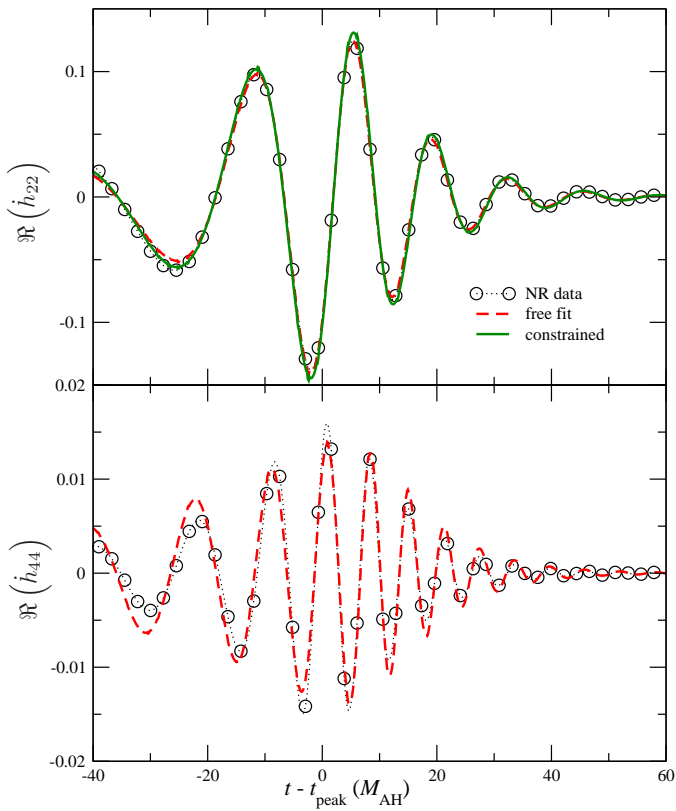}
\caption{Real part of the $(2,2)$ (top panel) and $(4,4)$ (bottom panel) strain-rate modes for
the \texttt{X1\_DD} run (extrapolated to $\Rext \rightarrow \infty$). Again, the
numerical data are indicated by circles, with a free fit over the parameters represented by the
dashed line. For the  $(2,2)$ mode, we also include the result of the fully restricted
waveform (see Eqs.~\eqref{eq:IRS_params_const_22}) as a continuous line.}
\label{fig:rehdot_fits_22_44_X1_DD}
\end{figure}

To contrast the quality of the fit performance for the $(2,2)$ and $(4,4)$ modes, we present in
Fig.~\ref{fig:rehdot_fits_22_44_X1_DD} the associated strain-rate (real parts) over the range of the
fit. We also plot in the upper panel the $(2,2)$ mode resulting from a fully constrained
model for amplitude and phase, using Eqs.~\eqref{eq:IRS_params_const_22}-\eqref{eq:IRS_A0_const_22}.

\section{Faithfulness of the Frequency Model}
\label{sec:match}

One way to quantify how much of the merger information we have captured by the 
modeling above is to compare the results of the model with the original fully
numerical waveforms in a detector context. We consider an explicit waveform model 
restricted to the $(2,\pm 2)$ modes, which contain most of the power. The waveform
phase is derived from integrating the model IRS frequency given by \eqref{eq:OmegaModel1}
and \eqref{eq:OmegaModel}, while the amplitude is given by \eqref{eq:AmpModel22_strainrate_trunc}.

In total there are seven parameters in these expressions, and one additional parameter
$\varphi_0$ arises as an integration constant in deriving the phase from our frequency model.  
Drawing on the results
of Sec.~\ref{ssec:constrained_model}, five of these parameters $\{c,\kappa,b,\Of,\alpha_1\}$
are specified by \eqref{eq:IRS_params_const_22} as functions of the final black hole's
leading quasinormal-mode frequency, thus reducing these free parameters to functions of
the final black hole's mass and spin.  Beyond these the only remaining parameters are
$\{\varphi_0,t_0,A_0\}$, corresponding to phase and time references, and an overall amplitude
scale.
For the
\texttt{X1\_DD} data, we can see the resulting waveform in the top panel of
Fig.~\ref{fig:rehdot_fits_22_44_X1_DD}.  For comparison we also show the results
of a ``free'' fit, where the frequency was fit to the NR data without the constraints in 
\eqref{eq:IRS_params_const_22}, and the amplitude was fit based on those results without
constraint on $\alpha_1$.  This is not a completely free fit, since the frequency parameters
are fit without regard for the consequences on the amplitude.

We calculate the mismatch for the
Advanced LIGO detector \cite{advligo_web}. ``Mismatch'' here is defined as the deviation of
the normalized overlap integral from unity, usually optimized over free parameters such as
overall phase and arrival time \cite{Owen:1995tm}:
\begin{equation}
\mbox{mismatch} \equiv 1 - \max_{\lambda_i} \frac{\langle h_m(\lambda_i) | h_e \rangle}{\sqrt{\langle h_m(\lambda_i) | h_m(\lambda_i) \rangle \langle h_e | h_e \rangle  }},
\label{eq:mismatch_def}
\end{equation}
where the frequency-space inner product $\langle \cdot | \cdot \rangle$ is the \emph{overlap}
between two signals, defined as \cite{Cutler:1994ys}
\begin{eqnarray}
\langle h_1 | h_2 \rangle &\equiv& 2 \int_{0}^{\infty} \frac{\left[\tilde{h}_1(f) \tilde{h}_2(f)^* + \tilde{h}_1(f)^* \tilde{h}_2(f)\right]}{S_n(f)} df \nonumber \\
                          &=& 4 {\rm Re} \left[ \int_{0}^{\infty} \frac{\tilde{h}_1(f) \tilde{h}_2(f)^*}{S_n(f)} df\right],
\label{eq:overlap_def}
\end{eqnarray}
where $\tilde{h}_1(f)$ and $\tilde{h}_2(f)$ are the Fourier transforms of the signals, and
$S_n(f)$ is the (one-sided) noise power spectral density of the detector we're interested in;
in this case, we take the ideal form $S_n(f)$ for Advanced LIGO given in Appendix A of
\cite{Reisswig:2009vc}:
\bea
S_n(f) &=& S_0 \left\{ x^{-4.14} - 5 x^{-2} \right. \nonumber \\
       & & \left. + 111 \left( 1 - x^2 +\frac{x^4}{2} \right) \left( 1 + \frac{x^2}{2} \right)^{-1}\right\},
\label{eq:Snf_AdvLIGO}
\eea
where $x \equiv f/f_0$, $S_0 = 10^{-49}$ and $f_0 = 215$ Hz.

\begin{figure}
\includegraphics*[width=3.5in]{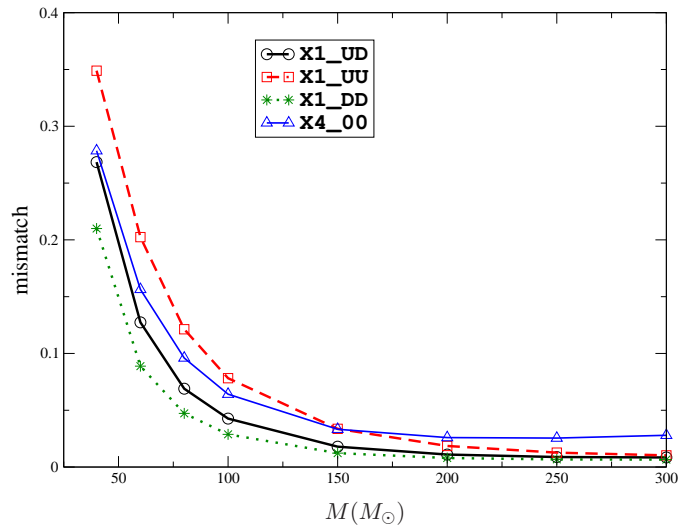}
\caption{Mismatch \eqref{eq:mismatch_def} between quadrupole $(2,\pm2)$ constrained IRS
and numerical-relativity waveforms for all simulations in the context of the Advanced LIGO
detector, where the system was observed along the polar axis at a distance of 1Gpc.}
\label{fig:mismatch_allX_quadNR}
\end{figure}

For this test, we pick black-hole binaries of total mass $M \in \{40\MSun,300\MSun\}$,
at a fixed distance of 1Gpc from the detector. Furthermore, we assume the system is
observed along the polar axis. The resulting mismatch with the quadrupole NR signal is shown for
all the runs (\texttt{X1\_00}, \texttt{X1\_UU}, \texttt{X1\_DD}, and \texttt{X4\_00}) in
Fig.~\ref{fig:mismatch_allX_quadNR}. Common to all cases is a sharp
falloff (that is, improvement) in mismatch as the system mass increases: from
$\gtrsim 25\%$ at $M = 40\MSun$ to $\lesssim 1\%$ for $M > 200\MSun$. This trend is
expected: overall physical frequencies scale inversely with system mass, so while the last
few pre-merger orbits' worth of radiation for a $40\MSun$ system might fall in Advanced LIGO's
most sensitive frequency band, only the higher-frequency merger and ringdown might lie in the
same band for a $200\MSun$ system.

\begin{figure}
\includegraphics*[width=3.5in]{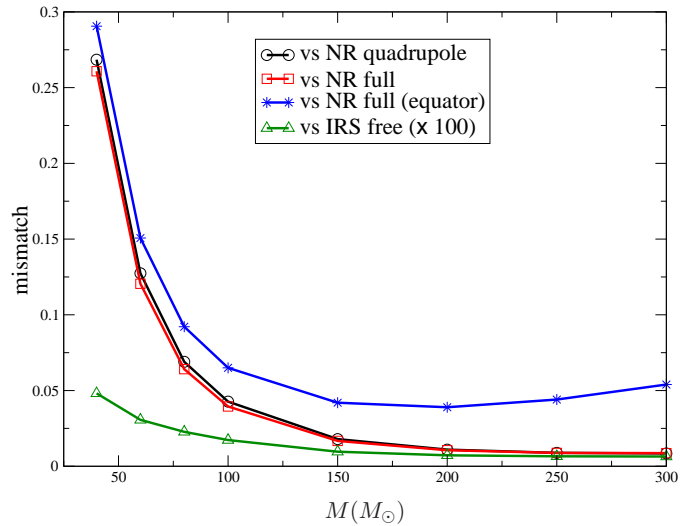}
\caption{Mismatch \eqref{eq:mismatch_def} for the \texttt{X1\_00} configuration, between
quadrupole $(2,\pm2)$ constrained IRS waveforms and (a) quadrupole numerical-relativity (NR)
waveforms along the polar axis (circles), (b) full NR waveforms along the polar axis (squares),
(c) full NR waveforms on the equator (stars), and (d) the unconstrained IRS waveforms along the
polar axis (triangles). All mismatches were calculated in the context of the Advanced LIGO
detector, at a distance of 1Gpc.}
\label{fig:mismatch_X1_00}
\end{figure}

In Fig.~\ref{fig:mismatch_X1_00} we again show this mismatch for the \texttt{X1\_00} configuration,
along with three other mismatches: with the \emph{full}
NR signal (all modes) along the polar axis, and on the equatorial plane, and also against
the \emph{unconstrained} model, using ``free''-fit parameters (Table~\ref{table:fit_params_free}).
This last mismatch has been scaled up by a factor of 100 for visibility.

For these results we have compared with just the $(2,2)$ component of our numerical simulation 
waveforms.  If applied in an actual observation there would be additional power, perhaps at the
level of up to several percent in other harmonics.  This model makes no attempt to fit those
contributions.  With a little work we could extend our constrained model to approximate the
contributions of these other modes, but this would necessarily require dependence on many 
additional parameters, including information about the component masses and spins and the 
relative orientation of source and detector.

We note that the the ``sweet spot'' of the Advanced LIGO sensitivity curve \eqref{eq:Snf_AdvLIGO}
is around 200 Hz. For systems at the low-mass end of our plot, $40 \MSun$, this frequency range is
accurately handled by post-Newtonian-based waveforms. Thus a full waveform model appropriate for
such low-mass systems should really be a combination of our merger-ringdown model with a PN-based inspiral.
By neglecting this here, and integrating over the full band, we will suffer from junk numerical
radiation and windowing artifacts at the lower-mass end. It is reasonable to suspect that we would get a
considerably lower mismatch if we restricted our integration to frequencies above
$M\omega_{22} \approx 0.15$.  The estimates presented in
Figs.~\ref{fig:mismatch_allX_quadNR}-\ref{fig:mismatch_X1_00} are therefore likely to be conservative.

With one detector, signals based on the waveform model we have constructed here depend only on 
the intrinsic parameters describing the final black hole, phase and time references and overall amplitude,
just five of the 17 parameters describing generic black-hole merger observations.  Our results suggest that
dominant features of the powerful merger-ringdown radiation may be described with little or no reference to the details of
the component black holes. This provides a complementary description of the merger to those based on the binary
inspiral parameters.
Even without a long inspiral lead-in, such models may be useful in detecting gravitational waves from high-mass mergers.
Our waveforms are based on the same parameters as those in ringdown-based approaches to merger observations in ground-based
detector pipelines \cite{LSC-Virgo_GWDA,Goggin:2008dz,Cadonati:2009jg,Sengupta:2009nm} and may be useful in future 
versions of these searches.

\section{Discussion}
\label{sec:discuss}

In this paper, we have investigated the mode-decomposed gravitational waveforms resulting from
the merger of aligned-spin black-hole binaries. Our primary purpose was to establish how
well the implicit rotating source (IRS) picture of the binary as a GW source -- first suggested
in \cite{Baker:2008mj} in the context of nonspinning holes of comparable mass ratios -- holds
in this different branch of parameter space.

Based on these investigations, we note that the modal structure of aligned-spin mergers is like that
of the equal-mass nonspinning configuration, dominated by  the $(2,\pm 2)$, $(3,\pm 2)$,
and $(4,\pm 4)$ modes.  These modes still display IRS-type behavior, featuring common rotational 
phase evolution with little offset through late-inspiral, merger, and into ringdown.
The peak modal amplitudes are similar to those for the nonspinning case, though the duration of the
peak region (which was roughly independent of mass ratio) is extended for aligned spins 
(and shortened for anti-aligned spins). A similar timescale dependence is seen in the rise to peak
frequency. 

In applying our late-merger frequency model (with a slightly modified parametrization) to these
new cases, we have found that the model still performs well for the dominant modes.
We enhance the original mode amplitude model of \cite{Baker:2008mj} to achieve improved behavior, at
least for the leading $(2,\pm 2)$ modes; however it yields up to $\sim 10\%$ mismatches at and before
peak for the next most important modes.

For the $(2,\pm 2)$ modes at least, we have attempted to constrain all parameters explicitly with 
reference only to the state of the final black hole (i.e., its dominant quasinormal-mode complex frequency).
With these constraints, we have reduced the additional free parameters to just three: $t_0$, the time of 
maximum chirp-rate, $\varphi_0$, the phase offset, and $A_0$, the mode's amplitude scale.  
This description provides an approximate fit to the late part of the waveforms for all our simulations, 
including equal-mass spinning cases, and the 4:1 nonspinning case.  Moving back in time to earlier points
before the merger, the quality of this fit degrades and other physical details of the premerger binary
become more significant.  We see evidence of this when comparing the \texttt{X4\_00} (4:1 nonspinning)
and the \texttt{X1\_DD} (equal-mass down-down spins) configurations, which result in the same final spin.

We have quantified the quality of this approximation by calculating Advanced LIGO fitting factors.
For system masses of
$\gtrsim 150 \MSun$, we have found mismatches of $\lesssim 5\%$ between the full numerical-relativity
waveform and the $(2,\pm 2)$-mode-only model waveforms.

Our results suggest that an approach to gravitational-wave observation templates with parameters 
tied first to the structure of the final black hole may be useful for Advanced LIGO observations  
of intermediate-mass mergers.  
These would be an alternative
to the time-domain effective-one-body (EOB) templates of \cite{Pan:2007nw,Pan:2009wj,Pan:2010hz}, and
the frequency-domain ``phenomenological'' templates of \cite{Ajith:2009bn,Santamaria:2010yb}, similar
to ringdown searches currently being applied to LIGO data \cite{LSC-Virgo_GWDA,Goggin:2008dz,Cadonati:2009jg,Sengupta:2009nm}.  In future work
we plan to investigate the quality of this model, or its extensions, for a broader variety of mergers,
including precessing and eccentric configurations.

In this explicit modeling, we have focused on the dominant $(2,2)$ mode. 
A similar model incorporating full multi-mode information can also be applied to complete inspiral-merger-ringdown waveform
templates, as was done for nonspinning systems in \cite{Baker:2008mj}. 
Future work on this topic will focus on improving and extending the amplitude model to cover multiple
significant modes of the merging binary.

\acknowledgments

This work was supported by NASA grants 08-ATFP08-0126 and 09-ATP09-0136.
Resources supporting this work were provided by the NASA High-End Computing (HEC) Program through
the NASA Advanced Supercomputing (NAS) Division at Ames Research Center.


\appendix*

\section{Convergence}
\label{appendix:convergence}

In this appendix, we present the convergence properties of the
evolution fields and extracted waveforms. Our presentation will model
that of \cite{Baker:2008mj}.

The simulations \texttt{X1\_00}, \texttt{X1\_DD}, and \texttt{X1\_UD} used
identical numerical methods and grid structures, with finest (near-puncture)
resolutions of $3M/128$, $3M/160$, and $M/64$, and wave-extraction-region
resolutions of $3M/2$, $6M/5$, and $M$, respectively. The \texttt{X1\_UD} simulation
was also carried out at an ultra-high resolution of $3M/224$ (wave-extraction
resolution $6M/7$).

The remaining equal-mass simulation, \texttt{X1\_UU}, uses identical numerical
methods at the same resolutions, but had a different grid structure in the
wave-extraction zone. This has been seen to result in higher noise levels in
waveform quantities, but should not affect the overall convergence properties.
We will use the \texttt{X1\_UD} resolution to assess convergence levels for the
{\sc Hahndol/Paramesh} code. 

\subsection{Constraints}

To establish constraint convergence in the equal-mass runs, we look at the L1-norms
of the Hamiltonian and momentum constraints both in the strong-field region and
in the region where the waveforms are extracted.

\begin{figure}
\includegraphics*[width=3.0in]{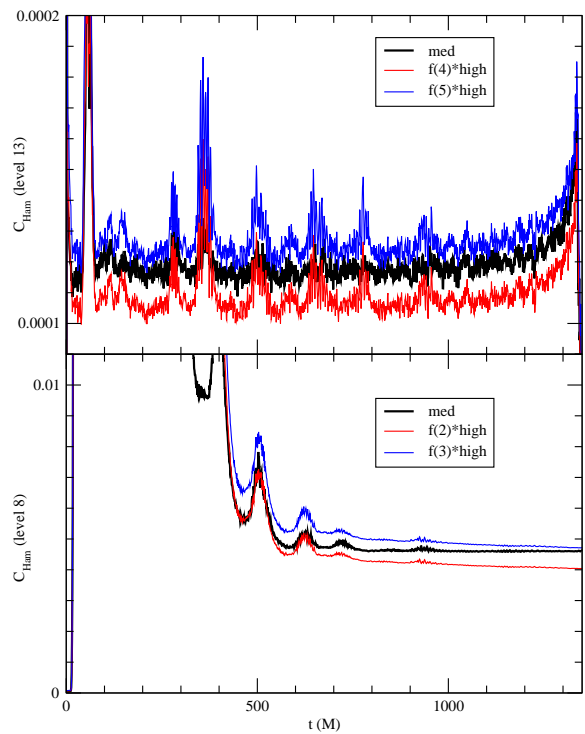}
\caption{Convergence of the Hamiltonian constraint's L1-norm for the \texttt{X1\_UD}
simulation: between fourth- and fifth-order convergence in level 13 (upper panel); between
second- and third-order convergence in level 8 (lower panel).}
\label{fig:CHam_conv_X1_UD}
\end{figure}

Figure~\ref{fig:CHam_conv_X1_UD} shows the L1-norm of the Hamiltonian constraint.
In the upper panel, level 13 (the region just outside the punctures) demonstrates
between fourth- and fifth-order convergence; in the lower panel, level 8 (containing the
wave-extraction spheres) demonstrates diminished convergence -- between second and
third order. Note that the resolution in these outer regions is much lower than in the
crucial high-resolution regions, where the black holes reside. It is our
understanding that errors in these distant regions are dominated by the effects of
uninteresting short-wavelength features (particularly gauge modes), which propagate 
out from the center and become poorly resolved in the coarse regions.

\begin{figure}
\includegraphics*[width=3.0in]{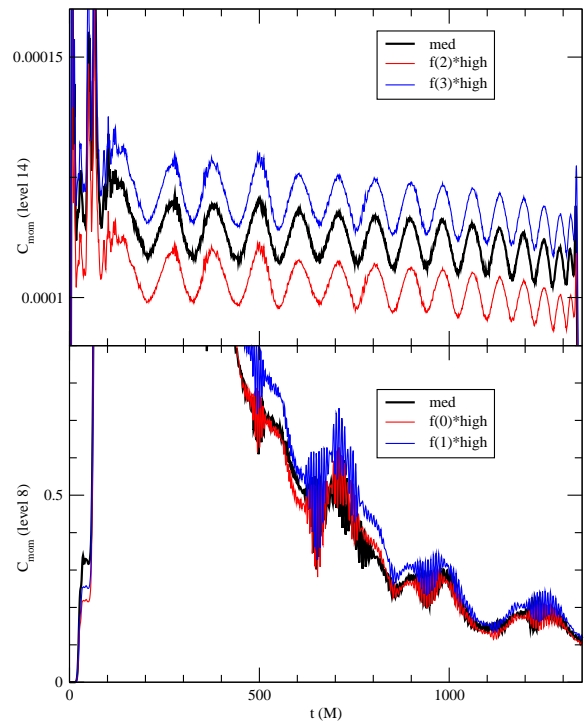}
\caption{Convergence of the momentum constraint's L1-norm for the \texttt{X1\_UD}
simulation: between second- and third-order convergence in level 14 (upper panel); without
any established convergence order in level 8 (lower panel).}
\label{fig:Cmom_conv_X1_UD}
\end{figure}

Convergence is more difficult to establish for the momentum constraint.
Figure~\ref{fig:Cmom_conv_X1_UD} shows the behavior of the $x$ component's L1-norm in
level 14 (the region containing the punctures), and level 8 (containing the
wave-extraction spheres).
The very inner zone (level 14) -- displays clean behavior: all components of the momentum
constraint are roughly 2.6-order convergent. Once we move outside this finest region, however,
convergence drops down precipitously.
We note, however, that the momentum constraint in general is two to three orders of magnitude
lower than the Hamiltonian constraint, which suggests that there is a small amount of
low-order error present in all constraints, but which is dominated by higher-amplitude
(but convergent) error only in the Hamiltonian constraint.

{\sout Constraint violation information was not available for the Cactus-based \texttt{X4\_00}
simulations.}

\subsection{Waveforms}

\begin{figure}
\includegraphics*[width=3.0in]{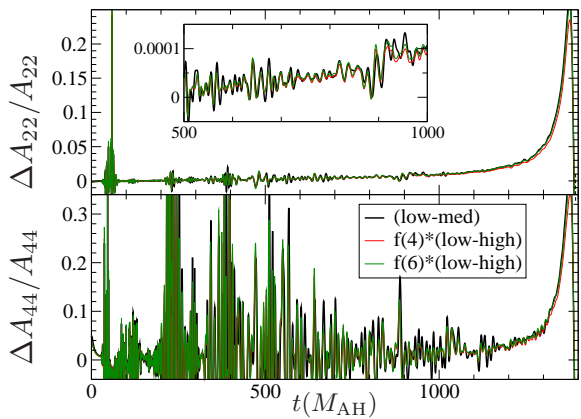}
\caption{Amplitude convergence for the $(2,2)$ (upper panel) and $(4,4)$ (lower
panel) modes of the \texttt{X1\_UD} simulations, based on the central resolutions
$3M/160$, $M/64$, and $3M/224$. In each case, the ($3M/160 - 3M/224$) difference has
been scaled up assuming fourth- and sixth-order convergence. Though noisy, the amplitude
differences appear to be consistent with sixth-order convergence throughout the inspiral,
until $\sim 150M$ before merger.}
\label{fig:amp_conv_X1_UD}
\end{figure}

\begin{figure}
\includegraphics*[width=3.0in]{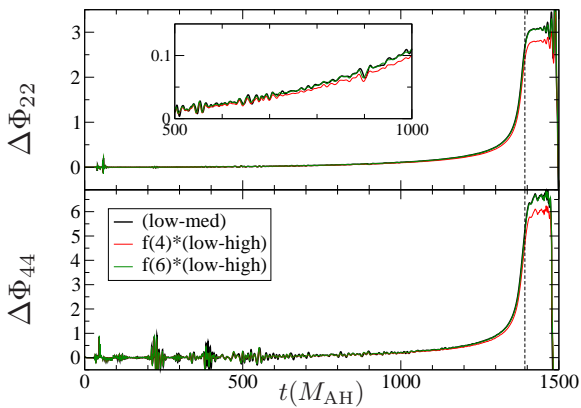}
\caption{Phase convergence for the $(2,2)$ (upper panel) and $(4,4)$ (lower
panel) modes of the \texttt{X1\_UD} simulations, based on the central resolutions
$3M/160$, $M/64$, and $3M/224$. In each case, the ($M/64 - 3M/224$) difference has
been scaled up assuming fourth- and sixth-order convergence. The phase differences
appear to be consistent with sixth-order convergence throughout the inspiral, merger,
and ringdown.}
\label{fig:phase_conv_X1_UD}
\end{figure}

In Figs.~\ref{fig:amp_conv_X1_UD} and \ref{fig:phase_conv_X1_UD}, we demonstrate
sixth-order convergence for the $(2,2)$ and $(4,4)$ modes' amplitudes and phases,
respectively.

\begin{figure}
\includegraphics*[width=3.0in]{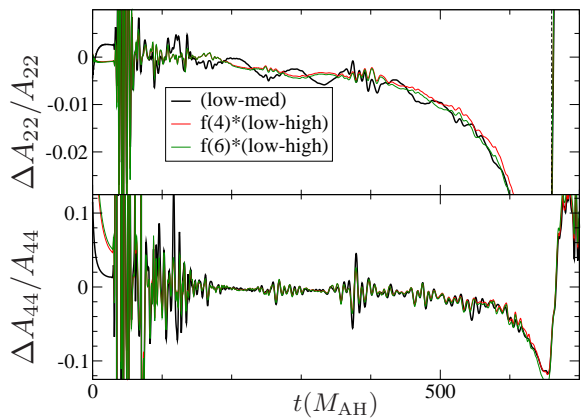}
\caption{Amplitude convergence for the $(2,2)$ (upper panel) and $(4,4)$ (lower
panel) modes of the \texttt{X4\_00} simulations, based on the central resolutions
$M/96$, $M/128$, and $M/160$. In each case, the ($M/96 - M/160$) difference has
been scaled up assuming fourth- and sixth-order convergence. The amplitude differences
appear consistent with sixth-order convergence until approximately $60M$ before peak,
when the rate declines to close to fourth-order.}
\label{fig:amp_conv_X4_00}
\end{figure}

For the 4:1 data (Fig.~\ref{fig:amp_conv_X4_00}), we see generally cleaner waveforms,
but also a large oscillation in errors until about $400M$ into the evolution.
After this, convergence appears to be sixth-order until close to amplitude peak time,
when it declines to fourth-order. This may be because the overall error is dominated by
uncertainties in the merger time, which is determined by the fourth-order-accurate Runge-Kutta
time-integration scheme. Unfortunately, we could not disentangle this effect to sufficient
accuracy to establish sixth-order convergence through the peak time.

\begin{figure}
\includegraphics*[width=3.0in]{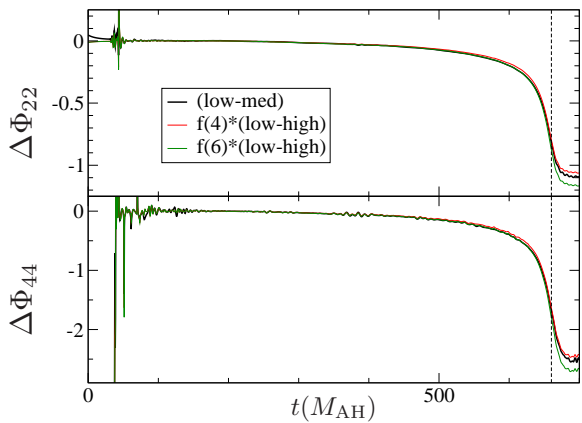}
\caption{Phase convergence for the $(2,2)$ (upper panel) and $(4,4)$ (lower
panel) modes of the \texttt{X1\_UD} simulations, based on the central resolutions
$M/96$, $M/128$, and $M/160$. In each case, the ($M/96 - M/160$) difference has
been scaled up assuming fourth- and sixth-order convergence. The phase differences
appear to be consistent with sixth-order convergence through late inspiral, declining
to closer to fifth-order at merger and ringdown.}
\label{fig:phase_conv_X4_00}
\end{figure}

The 4:1 waveform phase evolution (Fig.~\ref{fig:phase_conv_X4_00}) also seems to display
sixth-order convergence until the merger, when it declines to fifth-order. We note that
the scale of the errors is generally less than half those of the \texttt{X1\_UD} data from
Fig.~\ref{fig:phase_conv_X1_UD}.

\end{document}